\documentclass{aa}
\usepackage{natbib}
\usepackage{graphicx}
\usepackage{epstopdf}
\usepackage{txfonts}
\begin{document}

   \title{Temporal and spatial variations of the absolute reflectivity of Jupiter and Saturn from 0.38 to 1.7 $\mu$m with PlanetCam-UPV/EHU \thanks{Tables A1-A4 and B1-B8 are only available in electronic form at the CDS via anonymous ftp to cdsarc.u-strasbg.fr (130.79.128.5) or via http://cdsweb.u-strasbg.fr/cgi-bin/qcat?J/A+A/}}
\titlerunning{PlanetCam absolute photometry}

   \author{I. Mendikoa
          \inst{1},
          A. S\'{a}nchez-Lavega
          \inst{1},
          S. P\'{e}rez-Hoyos
          \inst{1},
          R. Hueso
          \inst{1},
          J.F. Rojas
          \inst{1},
          \and
          J. L\'{o}pez-Santiago
          \inst{2}
          }
          \authorrunning{Mendikoa et al.}

   \institute{Departamento de F\'{i}sica Aplicada, Escuela de Ingenier\'{i}a, Universidad del Pa\'{i}s Vasco, Plaza Torres Quevedo 1, E-48013 Bilbao, Spain\\
              \email{inigo.mendikoa@tecnalia.com}
         \and
             Department of Theory of Signal \& Communications, Universidad Carlos III de Madrid, Avenida de la Universidad 30, Legan\'{e}s, E-28911 Madrid, Spain\\
              \email{jalopezs@ing.uc3m.es}
             }

   \date{Received May 4, 2017; accepted July 22, 2017}


  \abstract
   {}
   {We provide measurements of the absolute reflectivity of Jupiter and Saturn along their central meridians in filters covering a wide range of visible and near-infrared wavelengths (from 0.38 to 1.7 $\mu$m) that are not often presented in the literature. We also give measurements of the geometric albedo of both planets and discuss the limb-darkening behavior and temporal variability of their reflectivity values for a period of four years (2012-2016).}
   {This work is based on observations with the PlanetCam-UPV/EHU instrument at the 1.23 m and 2.2 m telescopes in Calar Alto Observatory (Spain). The instrument simultaneously observes in two channels: visible (VIS; 0.38-1.0 $\mu$m) and short-wave infrared (SWIR; 1.0--1.7 $\mu$m). We obtained high-resolution observations via the lucky-imaging method.}
   {We show that our calibration is consistent with previous independent determinations of reflectivity values of these planets and, for future reference, provide new data extended in the wavelength range and in the time. Our results have an uncertainty in absolute calibration of 10--20\%. We show that under the hypothesis of constant geometric albedo, we are able to detect absolute reflectivity changes related to planetary temporal evolution of about 5-10\%.}
   {}


   \maketitle
%

\section{Introduction}

   Knowledge of the absolute reflectivity of planetary images is essential for precise modeling of the light scattered by planetary atmospheres. This is particularly important for Jupiter and Saturn, given the spatial and temporal variability of their atmospheres. The combination of observations at different wavelengths from the visible to the near-infrared allows for probing of a wide range of altitudes in the atmospheres of these planets but requires an absolute calibration of the images. The idea of using images of the giant planets acquired in different methane absorption bands to gain insight into the horizontal and vertical structure of their atmospheres is not new \citep{west1980,west1982,west1983b}, but acquiring enough multispectral images at high spatial resolutions is generally difficult. When using multispectral imaging, the loss of spectral resolution can be compensated by the gain in spatial resolution, particularly if filters are properly defined to match the absorption bands and adjacent continuums and other wavelengths of interest, such as the so-called chromophore absorptions at short wavelengths defining the red colors of Jupiter and pale tones of Saturn \citep{west2007, west2009}.

Although there are many works dealing with calibrated observations in the visible part of the spectrum for Jupiter and Saturn, including Hubble Space Telescope (HST) calibrated observations \citep{sph2005,karkoschka2005} or spacecraft imaging \citep{sph2016} and ground-based multispectral photometry \citep{mallama}, the near-infrared side of the spectrum (SWIR; short wavelength infrared) has been analyzed less often \citep{clark1979,depater2010}. This part of the spectrum is of particular interest to prepare for James Webb Space Telescope (JWST) image observations with the NIRCAM instrument of Jupiter to Neptune since JWST will achieve a spatial resolution better than Hubble Space Telescope (HST) from 0.7 to 2.1 microns \citep{norwood}. In this work, we profusely refer to a number of previous studies that served as an external reference for the absolute reflectivity values we obtain in this work. No seasonal variability has been reported in the atmosphere of Jupiter except for planetary-scale changes at particular belts and zones not linked to seasonal changes, \citep[see][]{asl1996, asl2008, asl2017, fletcher2011, sph2012}; however, Saturn is known to display hemispherical seasonal changes and  changing viewing or illumination geometry owing to the tilt of its rotation axis \citep{west2009}. For this reason, Saturn photometric values usually only serve as a snapshot for a particular season. For visual wavelengths (380-1000 nm, covered by our VIS channel), \cite{karkoschka1998} provided the giant planets geometric albedo through medium-resolution spectroscopy, while \cite{chanover1996} covered a wide visible wavelength range providing Jupiter absolute reflectivity values. A similar work for the near-infrared part of the spectrum was presented by \cite{vincent2005}.

More recently, and at the same time as the PlanetCam observations, visible absolute reflectivity data were available from HST for both Saturn (July 2015) and Jupiter from the Outer Planet Atmospheres Legacy program (OPAL\footnote{https://archive.stsci.edu/prepds/opal/}; February 2016). Because data in the SWIR channel are much scarcer, we use the calibrated data presented in \cite{depater2010} for Jupiter and disk-integrated spectra at medium resolution from \cite{clark1979} as a reference for both Jupiter and Saturn. This makes clear that an updated reference for the reflectivity of Jupiter
and Saturn in an ample wavelength range is of interest for future works dealing with calibrated images from these planets and other giant planets outside our solar system.

Moreover, the availability of Jupiter and Saturn calibrated images in a consistent set of filters and through a number of observation campaigns can also provide support to space missions in the study of their atmospheres, as occurred in other ground-based observations of Jupiter and Saturn at the time of Voyager 1 and 2 flybys \citep{bergstrahl1981, west1982, cochran1982}. In the case of Jupiter, the NASA Juno mission, in orbit around Jupiter since June 2016 \citep{asl2017,hueso2017b}, is currently receiving ground-based support. Ground-based observations provide information at low phase angles complementary to space missions observing the planet at geometries with higher phase angles, as Jupiter during the Cassini flyby in 2000 \citep{mayorga} and the near-to-end nominal mission at Saturn \citep{sph2016}.

Here we present a multiwavelength (0.38 -- 1.7 $\mu$m) study of the absolute reflectivity of Jupiter and Saturn using high-spatial-resolution, ground-based images obtained with the instrument PlanetCam-UPV/EHU \citep{asl2012,mendikoa2016}. We present a set of observations from 2012 to 2016, which has allowed the creation of a wide database of Jupiter and Saturn calibrated images in reflectivity. This database will hopefully be enlarged with further observations to be retrieved in the future. Radiative transfer modeling from the calibrated series (i.e., sets of images with all the filters) presented here is not included and will be reported elsewhere.

The organization of the paper is as follows. The observations performed so far are presented in section 2. In Sect. 3, we focus on a set of PlanetCam photometric images of Jupiter and Saturn for all the filters available in both VIS and SWIR channels and describe the calibration procedure. We present PlanetCam calibration results in section 4, followed by a review of the different sources of uncertainty. Finally, in section 5 we compare these results with available reference values and discuss the temporal evolution of some atmospheric features over the four year period of these observation campaigns.
   \begin{table*}
      \caption[]{Observation campaigns at Calar Alto and standard stars used.}
      \label{obs_table}
      \centering
      \begin{tabular}{c c c c c c c}
      \hline \hline
      Campaign & Telescope & Instrument & Channel(s) & Objects & Standards & Spec. Type \\
      \hline
      Jul 28/30, 2012 & 1.23m     & PC1  & VIS      & Jupiter/Saturn    & HD195034  & G5D \tablefootmark{d} \\
                      &           &       &          &                & HD192281        &  O5f \tablefootmark{a}\\
      Dec  2/3, 2012  & 1.23/2.2m &     PC1      & VIS      & Jupiter        & HD19445   & sdF \tablefootmark{a} \\
                      &           &       &          &                & HD28068 &  O5f \tablefootmark{a}\\
      Apr 17/20, 2013 & 1.23m     &     PC1      & VIS      & Jupiter/Saturn    & HD84937   & sdF \tablefootmark{a} \\
                      &           &       &          &                & HD217086          &  O7V \tablefootmark{a}\\
                      &           &       &          &                & HD93521   &  O9Vp \tablefootmark{a}\\
      May 24/28, 2013 & 1.23m     & PC1   & VIS      & Saturn         & BD+26-2606  &      sdF \tablefootmark{a}   \\
                      &           &       &          &                & Hiltner102  &  B0III \tablefootmark{a}\\
                      &           &       &          &                & HD217086          &  O7V \tablefootmark{a}\\
      Nov  21/24, 2013& 1.23m     & PC1   & VIS      & Jupiter        & HD93521     & O9Vp \tablefootmark{a} \\
                      &           &       &          &                & HD217086    & O7V \tablefootmark{a} \\
                      &           &       &          &                & HD84937     & sdF \tablefootmark{a} \\
      Dec 13/20, 2013 & 1.23m     & PC1   & VIS      & Jupiter        & HD19445     &  sdF \tablefootmark{a} \\
                      &           &       &          &                & HD93521     & O9Vp \tablefootmark{a} \\
      Apr 5/10, 2014  & 1.23/2.2m & PC1   & VIS      & Jupiter/Saturn & HIP40047    & O5p \tablefootmark{a} \\
                      &           &       &          &                & HD93521     & O9Vp \tablefootmark{a} \\
                      &           &       &          &                & HD84937     & sdF \tablefootmark{a} \\
      May 12/15, 2014 & 1.23m     & PC1   & VIS      & Jupiter/Saturn & BD+17-4708  & sdF \tablefootmark{a} \\
                      &           &       &          &                & HIP40047    & O5p \tablefootmark{a} \\
      Jul 21/25, 2014 & 2.2m      & PC1   & VIS      & Saturn         & HD217086    & O7V \tablefootmark{a} \\
      Dec 11/13, 2014 & 2.2m      & PC2   & VIS/SWIR & Jupiter        & HD26965     & K0.5V C \tablefootmark{a} \\
                      &           &       &          &                & HD219477    & G2 II-III \tablefootmark{b} \\
      Mar 2/4, 2015   & 1.23m     & PC2   & VIS/SWIR & Jupiter        & HD95128     &  G1V \tablefootmark{b} \\
                      &           &       &          &                & HD87822     & F4V \tablefootmark{b} \\
                      &           &       &          &                & HD108519    & F0V \tablefootmark{b} \\
      May 21/25, 2015 & 2.2m      & PC2   & VIS/SWIR & Jupiter/Saturn & HD95128     & G1V \tablefootmark{b} \\
      Jul 10/13, 2015 & 2.2m      & PC2   & VIS/SWIR & Saturn         & HD1160      &  A0 \tablefootmark{c} \\
                      &           &       &          &                & HD4656      & K4 IIIb \tablefootmark{c} \\
                      &           &       &          &                & HD4628      & K2V \tablefootmark{c} \\
                      &           &       &          &                & HD173638    & F1 II \tablefootmark{b} \\
                      &           &       &          &                & HD16139     & G7.5 IIIa \tablefootmark{b} \\
      Dec 28/30, 2015 & 2.2m      & PC2   & VIS/SWIR & Jupiter        & HD28068     & G1V C \tablefootmark{d} \\
                      &           &       &          &                & HD93521     & O9Vp \tablefootmark{a} \\
                      &           &       &          &                & HD10307     & G1V \tablefootmark{b} \\
                      &           &       &          &                & HD75732     & G8V \tablefootmark{b} \\
      Mar 3/7, 2016   & 2.2m      & PC2   & VIS/SWIR & Jupiter/Saturn & HD19445     &  sdF \tablefootmark{a} \\
                      &           &       &          &                & BD+33-2642  & B2IVp D \tablefootmark{a} \\
      May 16/20, 2016 & 2.2m      & PC2   & VIS/SWIR & Jupiter/Saturn & HD192281    &  O5f \tablefootmark{a} \\
                      &           &       &          &                & HD84937     & sdF \tablefootmark{a} \\
                      &           &       &          &                & HD93521     & O9Vp \tablefootmark{a} \\
                      &           &       &          &                & HD179821    & G5Ia C \tablefootmark{b} \\
                      &           &       &          &                & HD95128     & G1V \tablefootmark{b} \\
                      &           &       &          &                & HD75555     & F5 C \tablefootmark{b} \\
      Jul 1/5, 2016   & 2.2m      & PC2   & SWIR     & Jupiter/Saturn & HD102870    & F8.5IV-V \tablefootmark{b} \\
                      &           &       &          &                & HD100006    & K0III D \tablefootmark{b} \\
                      &           &       &          &                & HD124850    & F7III C \tablefootmark{b} \\
                      &           &       &          &                & HD165782    & K0Ia B \tablefootmark{b} \\
      \hline
      \end{tabular}
          \tablefoot{
      \tablefoottext{a}{Isaac Newton Group of telescopes spectrophotometric database}
      \tablefoottext{b}{IRTF spectral library}
      \tablefoottext{c}{SIMBAD database}
      \tablefoottext{d}{Solar Analog}
      }
   \end{table*}

\section{PlanetCam observations}

\subsection{PlanetCam instrument}

PlanetCam-UPV/EHU \citep{mendikoa2016} is an astronomical camera designed for high-resolution imaging of solar system targets \citep{asl2012}. The instrument uses the lucky-imaging technique. This consists of the acquisition in a short time of several hundredths to thousands of short exposures (10’s of milliseconds) selected by their quality and stacked into a single image \citep{law2006} for broadband filters, and longer exposures (i.e., a few seconds each and also stacked for building high signal-to-noise ratio images) for narrowband imaging. The final stacked images are processed to obtain high-resolution observations that largely remove the blur from atmospheric effects \citep{mendikoa2016}. The selection of the best 1-10\% of frames without adding any additional image processing results in high-quality images that retain the photometry of the object observed.

The instrument contains several filters selected for color, broadband continuum, methane absorption bands, and their adjacent narrow continuum wavelengths chosen because of their interest for planetary studies. The camera works in the wavelength range from 0.38 to 1.7 $\mu$m supporting scientific research in atmosphere dynamics and vertical cloud structure of solar system planets. The design comprises two channels that work simultaneously by means of a dichroic beam splitter \citep{asl2012}. First channel (visible; VIS) covers wavelengths below 1 $\mu$m and the second channel (short-wave infrared; SWIR) works above this value and up to 1.7 $\mu$m. PlanetCam images have been used so far for several works on Jupiter \citep{hueso2017b,asl2017}, Saturn \citep{asl2014,asl2016a}, Venus \citep{asl2016b}, and Neptune \citep{hueso2017a}.

\subsection{Observation campaigns}

We performed several Jupiter and/or Saturn observation campaigns since the commissioning of PlanetCam in 2012 at Calar Alto Observatory in Spain via both the 1.23 m and 2.2 m telescopes. Table~\ref{obs_table} provides details on these campaigns, including the telescope and PlanetCam configuration \citep{mendikoa2016}. The PlanetCam1 (PC1) configuration means that only the visible channel (VIS) was available at the time, while PlanetCam2 (PC2) means that both the visible and infrared (SWIR) channels were available. There are also some differences in the optics between both configurations, but they are of no interest here since both the planet and standard stars are always observed with the same configuration. Table~\ref{obs_table} also shows the standard stars used for absolute reflectivity calibration including their spectral type.

\begin{figure*}
\centering
\includegraphics[width=17cm]{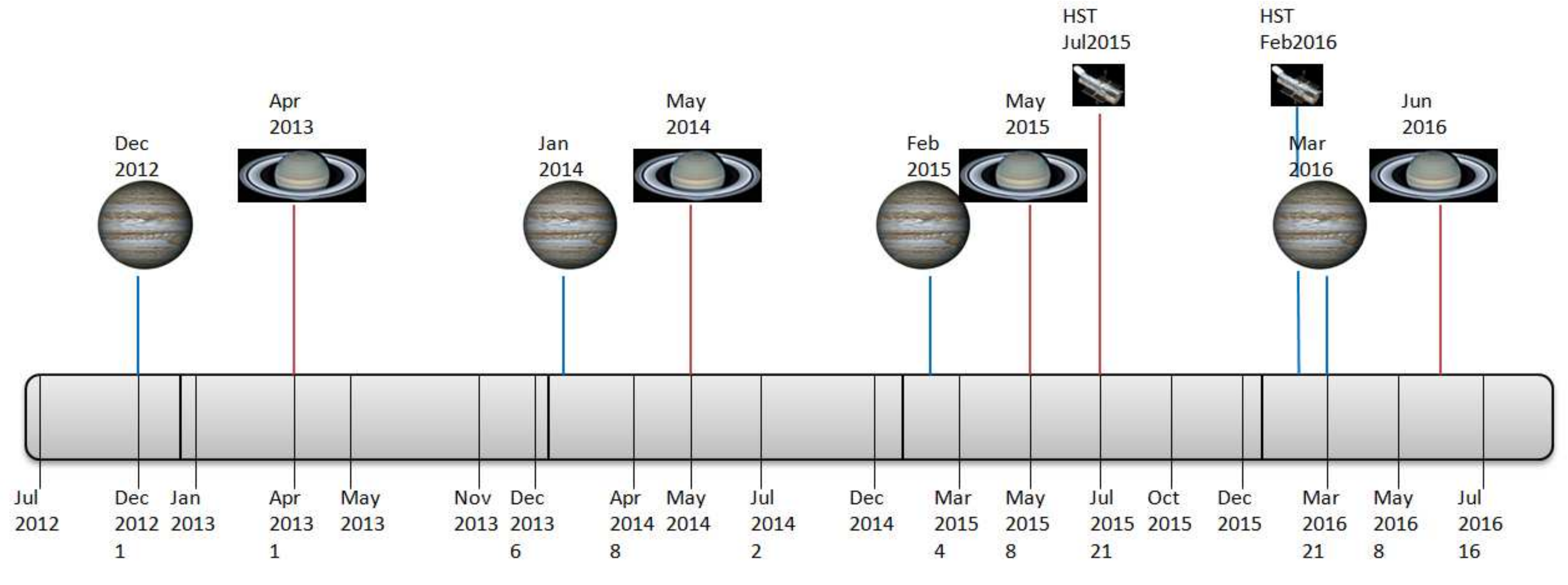}
\caption{Timeline of Jupiter and Saturn observations at Calar Alto. Images of the planets are shown at the opposition date, while HST indicates the availability of observations for cross-calibration. The number of calibrated series available at each date is shown below the graph.}
\label{fig_timeline}
\end{figure*}

A timeline representation of all these observation campaigns is shown in Figure~\ref{fig_timeline}, indicating Jupiter and Saturn opposition dates as well as HST relevant observations of Saturn (July 2015) and Jupiter (Feb 2016) used in this paper as reference values for  absolute reflectivity. The number of calibrated filter sequences is indicated below each campaign date.

As standard stars, we used a selection of stars from the Isaac Newton Group of telescopes spectrophotometric database\footnote{http://catserver.ing.iac.es/landscape/} for the VIS channel and from the IRTF spectral library\footnote{http://irtfweb.ifa.hawaii.edu/IRrefdata/sp\_catalogs.php} for the SWIR channel \citep{rayner2009}. These databases provide medium- to low-resolution spectra of the stars that can be convolved with our filters plus system response, thus getting accurate calibration factors as we explain later. Whenever this was not possible for all wavelengths, we completed the values using the VizieR photometry tool on the SIMBAD database\footnote{http://simbad.u-strasbg.fr/simbad/} \citep{ochsenbein2000}. The impact of such very low-resolution data for the standard stars that are discussed in section~\ref{uncertainty}.


The standard stars used for the absolute reflectivity calibration of Jupiter and Saturn were selected according to different criteria. First, the availability of a star with high-resolution spectrum was considered. Second, the position of the star in the sky relative to the planet that is calibrated was also considered, trying to get the standard star as close as possible to the planet with similar airmass at the corresponding observation time.

\section{Absolute reflectivity measurements}
\subsection{Photometric calibration}

In planetary sciences, it is common to use the absolute reflectivity $I/F$ of a given surface or atmosphere \citep{aslbook}. Omitting here the wavelength dependency, this quantity is defined as the ratio of surface brightness ($I$) to that of a flat Lambertian surface ($F$), where $\pi F$ is the solar flux (Wm$^{-2}$nm$^{-1}$) at the given planetary distance. A planet image can be converted from digital counts into absolute reflectivity $I/F$ with the intensity spectrum of a standard star as reference, from the expression below \citep{ortiz1993}:

  \begin{equation} \label{eq1}
      \left( \frac{I}{F} \right) _i = \frac{(C_p)_i}{\Sigma C_{star}} \frac{\int_{\lambda}\pi F_{star} \cdot T(\lambda)\Phi(\lambda)d\lambda}{\int_{\lambda}\pi F_{\odot} \cdot T(\lambda)\Phi(\lambda)d\lambda} D^2 \left(\frac{\pi}{\theta^2}\right) e^{\frac{k}{2.5}(X_p-X_{star})}
   .\end{equation}

This expression provides the absolute reflectivity $I/F$ at the pixel $i$ in the planet image. Here $(C_p)_i$ represents the count rate at the planet image pixel $i$, $\Sigma C_{star}$ the total number of count \textbf{rates} from the standard star, $\pi F_{star}$ the standard star flux outside atmosphere of the Earth (Wm$^{-2}$nm$^{-1}$) as retrieved from the database, $\pi F_{\odot}$ the solar flux at Earth (Wm$^{-2}$nm$^{-1}$) given by \cite{colina1996}, $D$ (in AU) is the distance from the Sun to the target at the observation time, $T(\lambda)$ the filter plus optics transmittance curve, $\Phi(\lambda)$ the detector sensitivity curve, and $\theta$ the image scale (arcsec/pixel) \citep{chanover1996}. The last term is added to consider the atmospheric effects, where $X_p$ and $X_{star}$ are the air masses for the planet and star, respectively, and $k$ the extinction coefficient of atmosphere of the Earth (magnitudes/air mass) at the time the images were taken.

\subsection{Image acquisition and processing}

\begin{figure}
\centering
\includegraphics[width=8cm]{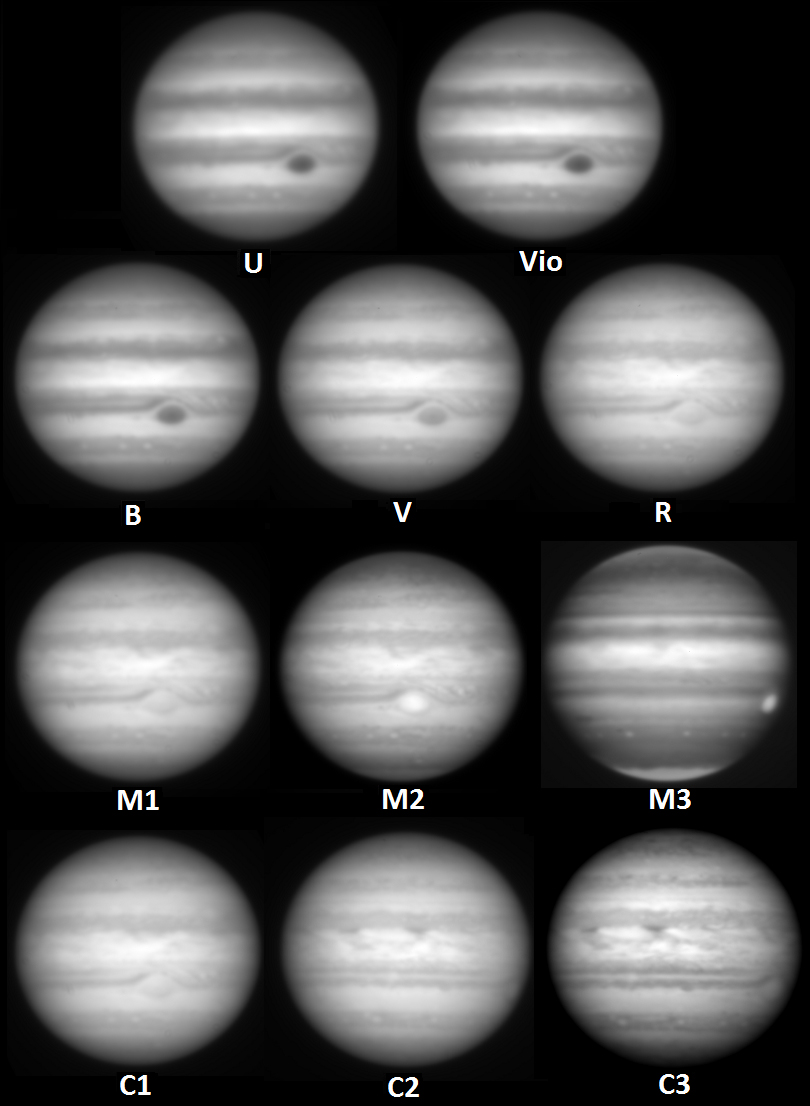}
\caption{Photometric images of Jupiter with PlanetCam VIS channel filters (March 3, 2016).}
\label{jupVIS}
\end{figure}

\begin{figure}
\centering
\includegraphics[width=8cm]{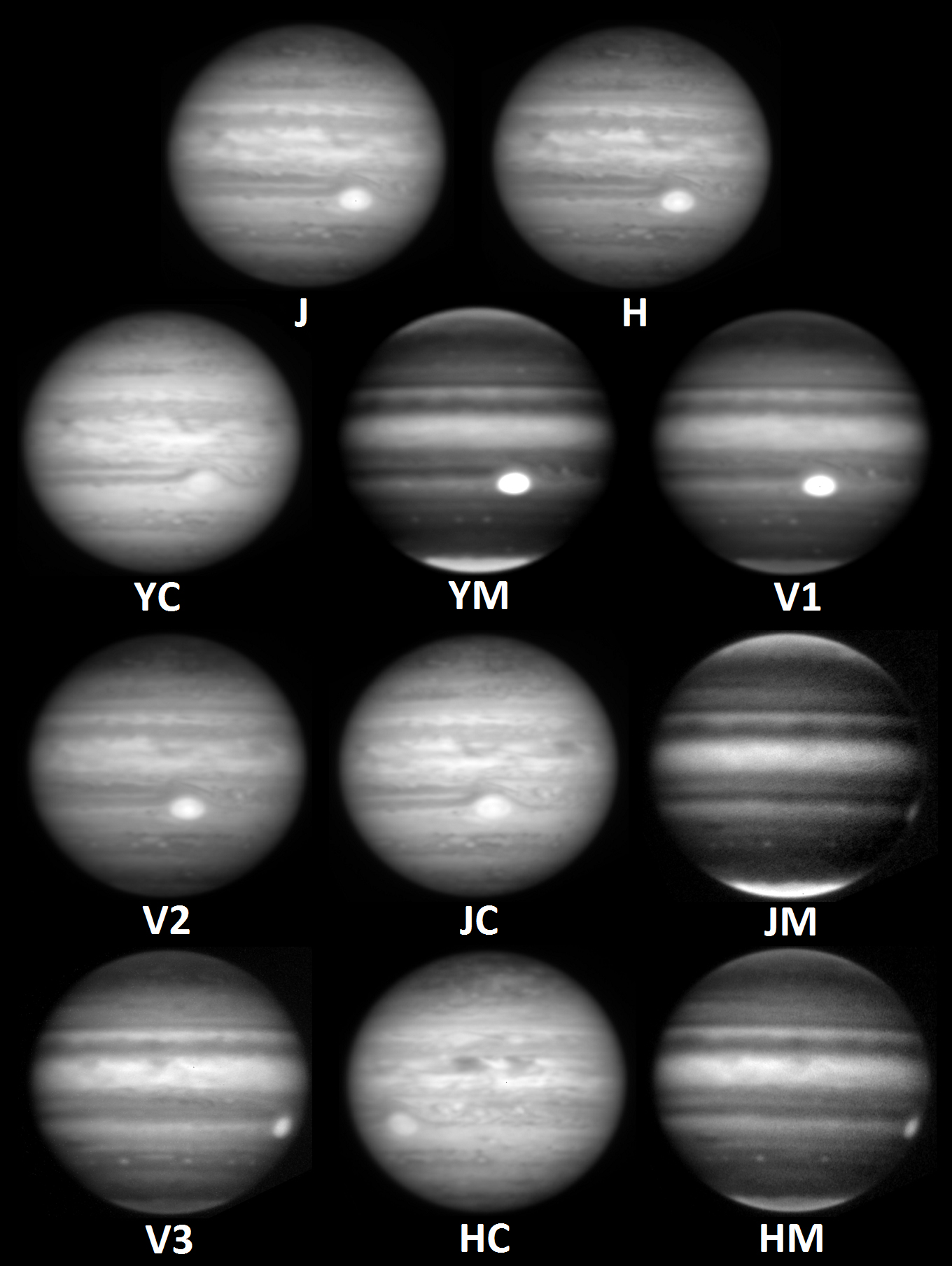}
\caption{Photometric images of Jupiter with PlanetCam SWIR channel filters (March 3, 2016).}
\label{jupSWIR}
\end{figure}

The calibration process begins with the capture of the sequence of frames of the science object (planet or calibration star) that are used to build the final image. This requires suitable exposure times for each filter to get a good signal-to-noise ratio that maintains detector linearity, while being fast enough to perform lucky-imaging at least for the wider filters. Images are processed with a pipeline written in IDL (Interactive Data Language) and specifically designed for PlanetCam. The  software, called PLAYLIST (PLAnetarY Lucky Images STacker), analyzes full directories of data locating for each science file the calibration files containing dark currents and flat fields (i.e., those matching exposure times and camera gain settings) to be used in the image reduction substracting the dark current and dividing by the flat field of each individual frame. Hot and cold pixels are removed then with an adaptive median filter technique. The quality of the resulting frame is analyzed by first finding the scientific object in the frame using a center of brightness algorithm that identifies the center of the planet (or star) and its size by defining a region of interest (ROI) of constant size through the sequence. Next, the quality of each frames is provided by a numeric metric based on the Sobel differential filter \citep{gonzalez2002}.  Sharp images have higher values of their spatial derivatives while unfocused or smooth images provide lower values. By summing up the absolute values of the Sobel filtered ROI containing the scientific target a good estimation of the frame quality is obtained. The software allows Lucy Richardson deconvolution \citep{richardson1972, lucy1974} on a frame-by-frame basis, which is used with bright images with low noise levels. Images are co-registered with a multiscale image correlation algorithm that matches images with a precision of one pixel. For each image sequence PLAYLIST generates several FITS files with different trade-offs between image quality in terms of spatial resolution and dynamic range. For instance, the best 1\% of all individual frames typically contain fewer atmospheric blurring but also low dynamic range, but the co-registered stack of all frames contains the maximum dynamic range but the highest level of atmospheric blurring. The software automatically produces selections with different percentages of the best frames (1\%, 2\%, 5\%, 10\%, 30\%, and all frames). Photometric images are saved in FITS files containing double precision real numbers representing the total digital counts per second for each pixel. The SWIR images are typically best represented by the images based on the best 1\%-5\% frames, while visible images in short wavelengths are generally better represented by the versions containing the best 10\% and very dark acquisitions in the strongest methane absorption filters generally require the versions based on the stack of all frames.

\begin{figure}
\centering
\includegraphics[width=8cm]{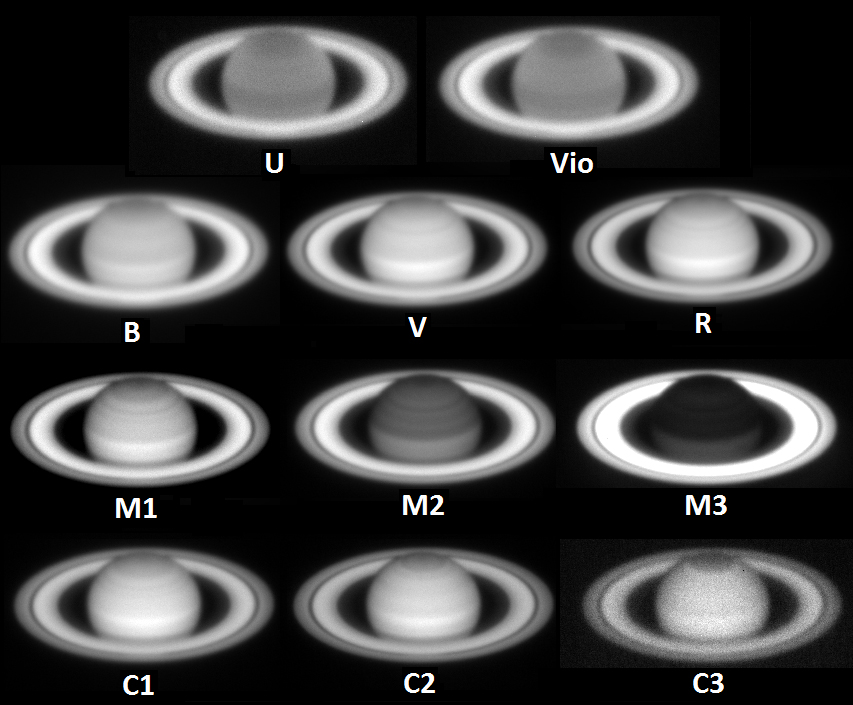}
\caption{Photometric images of Saturn in PlanetCam VIS channel filters (May 20, 2016).}
\label{satVIS}
\end{figure}

\begin{figure}
\centering
\includegraphics[width=8cm]{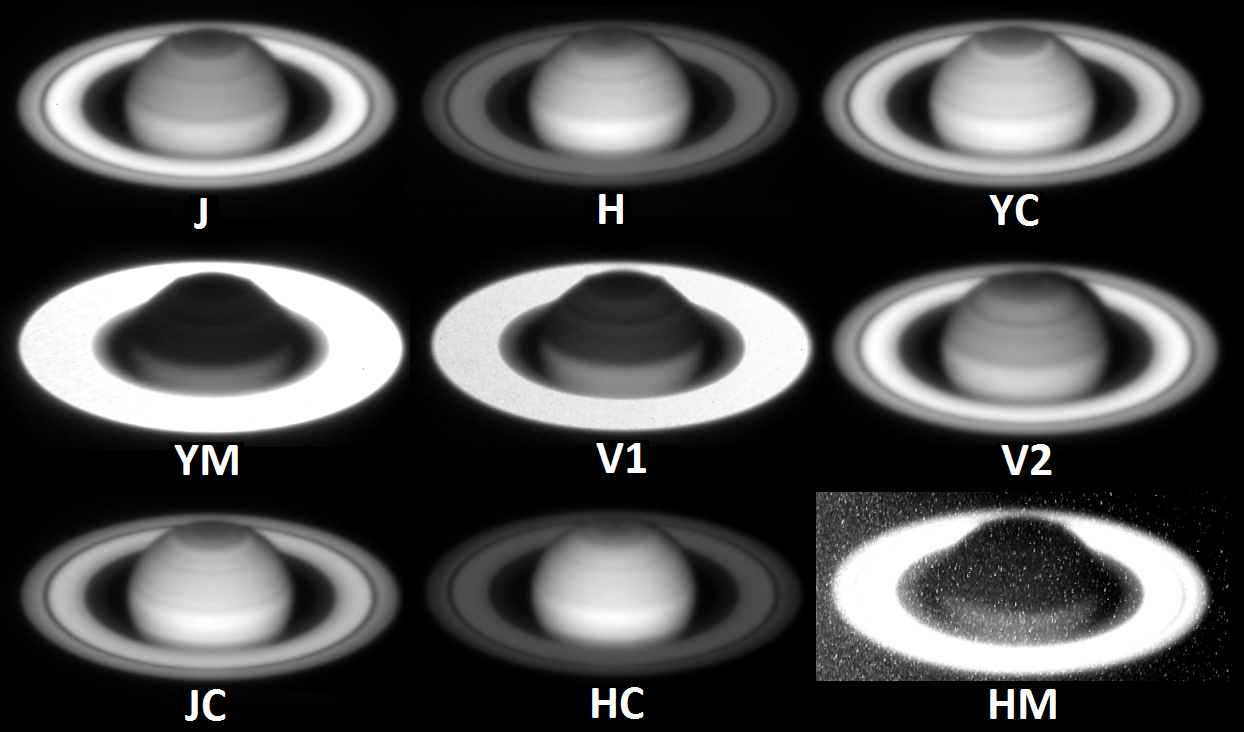}
\caption{Photometric images of Saturn in PlanetCam SWIR channel filters. The HM image is also affected by water absorption on the atmosphere of Earth and requires acquisitions with high gains resulting in a level of noise that cannot be fully subtracted from the dark current files (May 20th, 2016).}
\label{satSWIR}
\end{figure}

Figures~\ref{jupVIS} to~\ref{satSWIR} show photometric images of Jupiter and Saturn for different PlanetCam filters in both VIS and SWIR channels processed with PLAYLIST. These images preserve photometric information but can be high-pass filtered to show faint structures and study dynamics \citep[see, e.g.,][]{hueso2017b}. All the PlanetCam filters are listed by their short name and we provide below effective wavelength in nanometers for narrow filters \citep[full details can be found in][]{mendikoa2016}. The VIS channel includes filters U, Vio, B, V, R, M1 (619), C1 (635), M2 (727), C2 (750), M3 (890), and C3 (935), while SWIR channel includes YC (1090), YM (1160), V1 (1190), V2 (1220), JC (1275), JM (1375), V3 (1435), HC (1570), HM (1650), J, and H. Figure~\ref{jupVIS} shows photometric images of Jupiter taken with PlanetCam2 configuration at the 2.2 m telescope at Calar Alto observatory (March 2016) for absolute reflectivity calibration with VIS channel, while Figure~\ref{jupSWIR} shows those corresponding to SWIR channel. Figure~\ref{satVIS} shows photometric images of Saturn taken with PlanetCam2 at the 2.2 m telescope at Calar Alto observatory (May 2016) for absolute reflectivity calibration with VIS channel. Figure~\ref{satSWIR} shows Saturn images in the SWIR channel where image saturation at the rings in some filters does not affect the atmosphere photometric analysis.

\section{Results}

Following equation~\ref{eq1} we photometrically calibrated Jupiter and Saturn images, computing $I/F$ values for each pixel on the images. We show absolute reflectivity $I/F$ scans along central or sub-observer meridian (i.e., $I/F$ versus latitude) for all available filters at a resolution of 1$^{\circ}$ in latitude for each observation campaign, using different calibration standard stars detailed in table~\ref{obs_table}. Even though values at a fixed resolution of 1$^{\circ}$ are given, our photometric images without further processing have effective resolutions of 0.5--1'' which, in the case of Saturn, means that only 5--10$^{\circ}$ resolution is achieved. However, this also depends on the filter, with resolutions better that 0.5'' for wide filters used for lucky-imaging and overall better behavior in the SWIR channel \citep{mendikoa2016}.

Because of the zonal banding in the giant planets, this method is well suited to compare spatial and temporal reflectivity changes in different spectral regions. We obtain a mean $I/F$ scan for each filter and observation campaign. We computed inter-annual mean values from the yearly means from planetographic latitudes -80$^{\circ}$ to +80$^{\circ}$. Figures~\ref{jupVISif} to~\ref{satSWIRif} show the mean absolute reflectivity versus latitude at central meridian for each filter, including the standard deviation from the analysis of different image series in each campaign. The error sources are mainly due to image navigation and the uncertainties in the calibration, as we discuss in section~\ref{uncertainty}. We omitted the mean $I/F$ scans from some campaigns when the latitudinally averaged deviation from the overall mean $I/F$ scan was higher than the standard deviation at each latitude. After this filtering, we recomputed the overall mean $I/F$ scans and standard deviations, which resulted in the latitudinal scans shown in Figures~\ref{jupVISif} to~\ref{satSWIRif}. We provide a tabulated version of these data in supplementary tables A1 to A4, as they can be used as a reference for future works.

We used the empirical Minnaert law for diffuse reflection \citep{minnaert1941} to determine the $I/F$ variation with the cosines of the incidence ($\mu_0$) and emission ($\mu$) angles \citep{aslbook},

  \begin{equation} \label{eqminnaert}
      \left( \frac{I}{F} \right) (\mu , \mu_0) = \left( \frac{I}{F} \right)_0 \mu_0^k \mu^{k-1}
   ,\end{equation}

where $(I/F)_0$ represents the absolute reflectivity in absence of darkening effects at nadir viewing and $k$ is the limb-darkening coefficient. In this way images can be corrected from limb-darkening effects and photometric results from the central meridian can be extended to other longitudes. The mean values of Minnaert coefficients (from observations from 2012 to 2016) for each latitude and PlanetCam filter are presented in the form of tables in supplementary tables B1 to B8.

\subsection{Absolute reflectivity values}

\begin{figure}
\centering
\includegraphics[width=8cm]{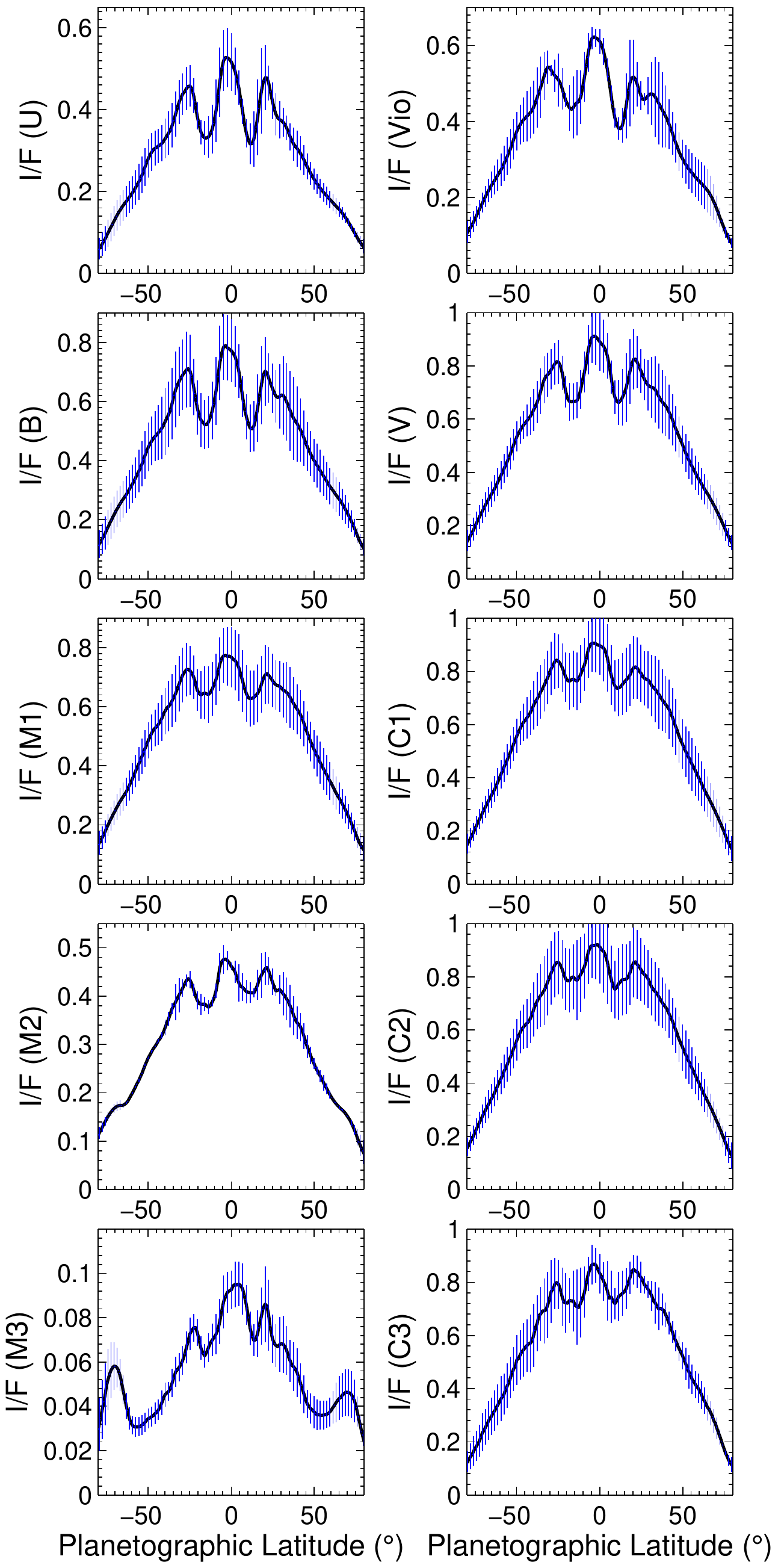}
\caption{North-south scans of the mean absolute reflectivity (2012--2016) of Jupiter across the central meridian in VIS channel filters.}
\label{jupVISif}
\end{figure}

\begin{figure}
\centering
\includegraphics[width=8cm]{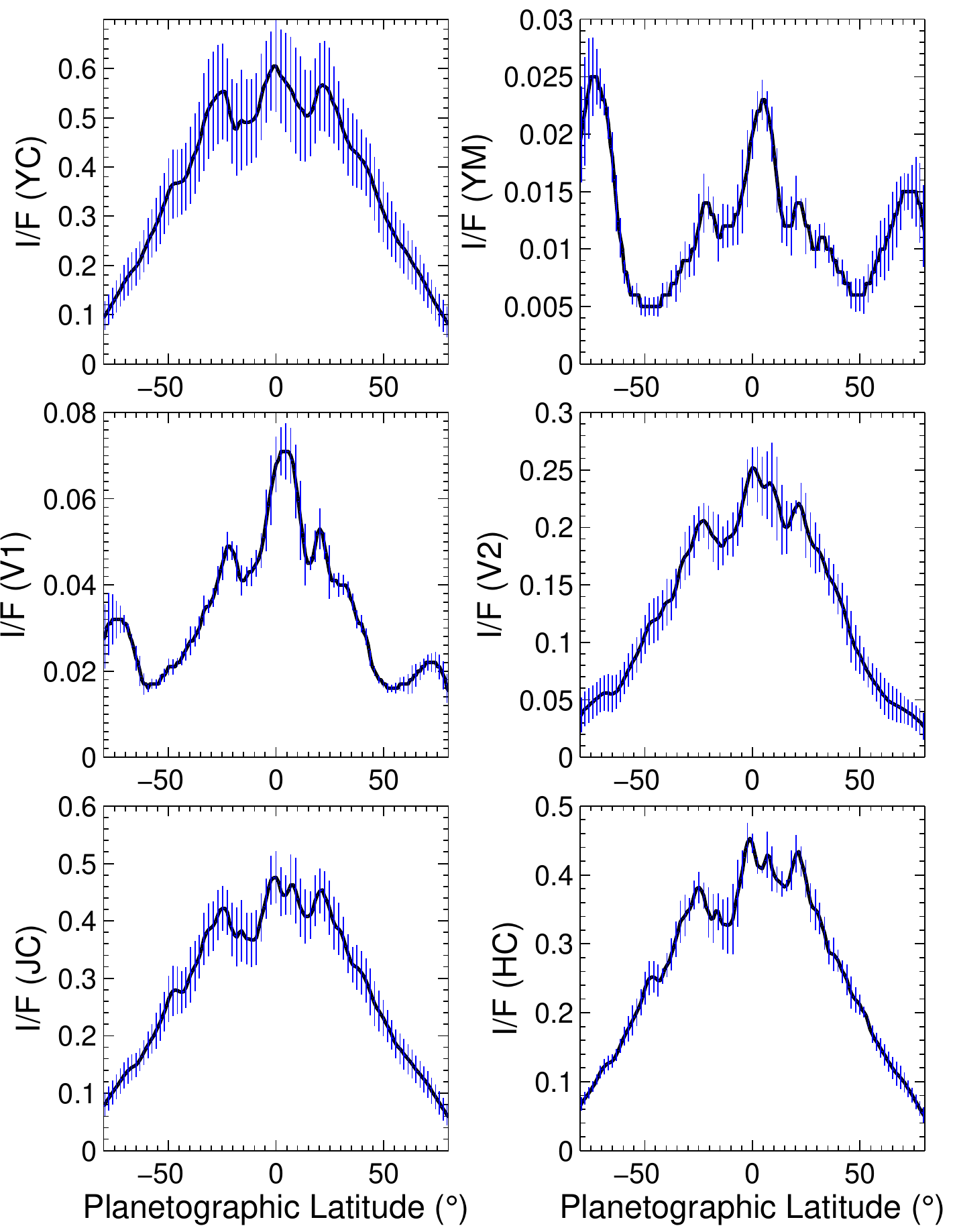}
\caption{North-south scans of mean absolute reflectivity (2015--2016) of Jupiter across the central meridian in SWIR channel filters.}
\label{jupSWIRif}
\end{figure}

Jupiter mean absolute reflectivity $I/F$ values along its central meridian (not using images with the Great Red Spot close to it) from observational campaigns in the visible channel between December 2012 and May 2016 (Figure~\ref{jupVISif}) are determined with different levels of uncertainty at different wavelengths. We calculated the latitudinal $I/F$ deviation from the mean for each filter from the ensemble of valid observations campaigns. For some filters only three campaigns provided valid data, while for others our sample grows to five campaigns. In the visible channel, the overall uncertainties considering the mean of standard deviations from all latitudes, for the filters represented in Fig.~\ref{jupVISif}, are as follows: UV (three campaigns; 16\%), Vio (three; 18\%), B (five; 23\%), V (four; 16\%), M1 (three; 20\%), C1 (three; 20\%), M2 (three; 11\%), C2 (three; 20\%), M3 (five; 18\%), and C3 (three; 18\%). These uncertainty values include, among different causes discussed in section~\ref{uncertainty}, possible physical changes due to temporal evolution, as discussed in section~\ref{temporal}.

In a similar way we show in Figure~\ref{jupSWIRif} the overall mean absolute reflectivity and associated uncertainty for SWIR filters corresponding to observations run between March 2015 and July 2016. Filters JM and V3 are not included in Fig. 7 because they are affected by telluric water absorption, while the HM filter is affected by the cutoff of the sensitivity of the PlanetCam SWIR detector (Mendikoa et al., 2016). The number of campaigns considered and average uncertainties for these filters are as follows: YC (four; 24\%), YM (four; 19\%), V1 (four; 12\%), V2 (three; 22\%), JC (four; 18\%), and HC (four; 11\%).

\begin{figure}
\centering
\includegraphics[width=8cm]{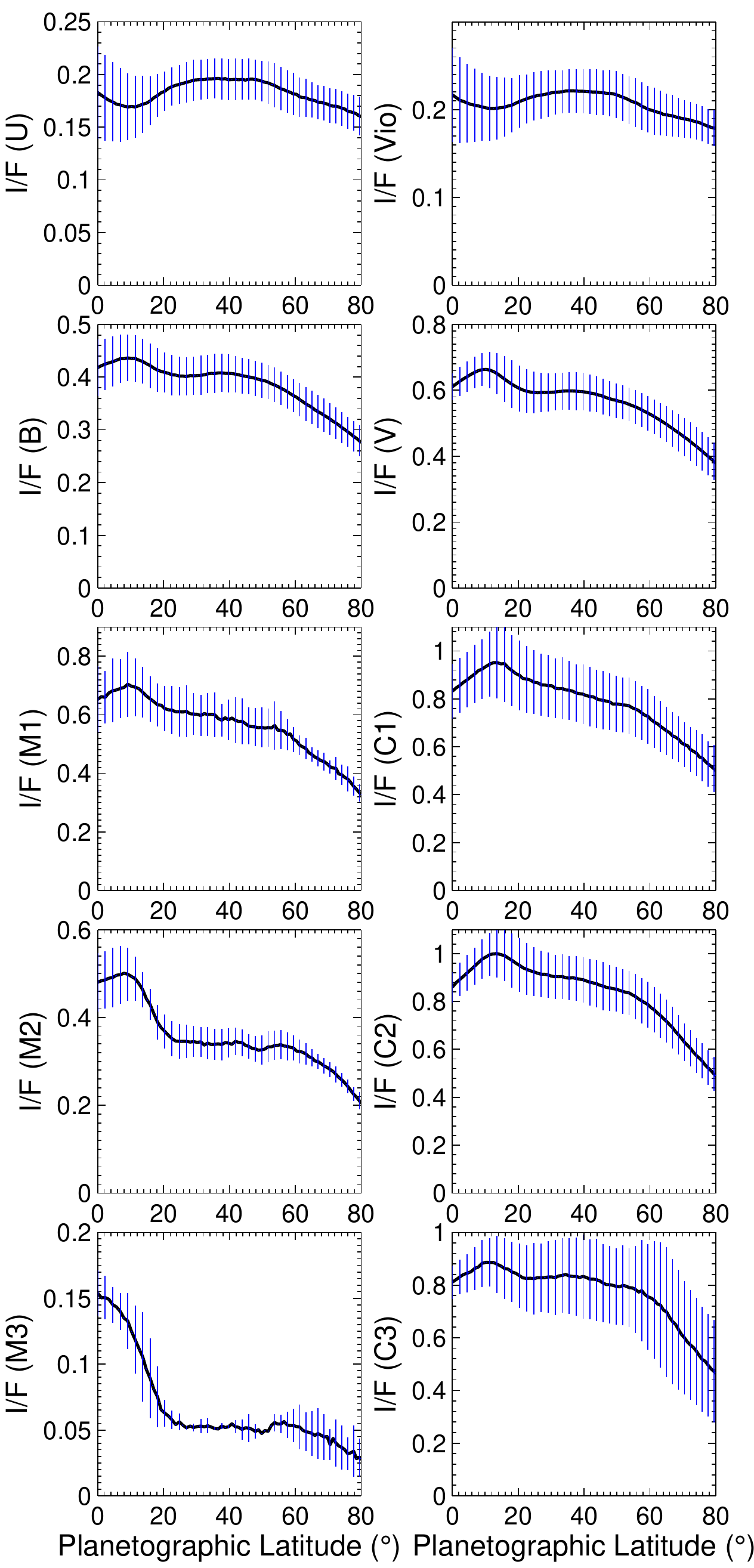}
\caption{North-south scans of mean absolute reflectivity (2013--2016) of Saturn across the central meridian in the northern hemisphere in VIS channel filters.}
\label{satVISif}
\end{figure}

\begin{figure}
\centering
\includegraphics[width=8cm]{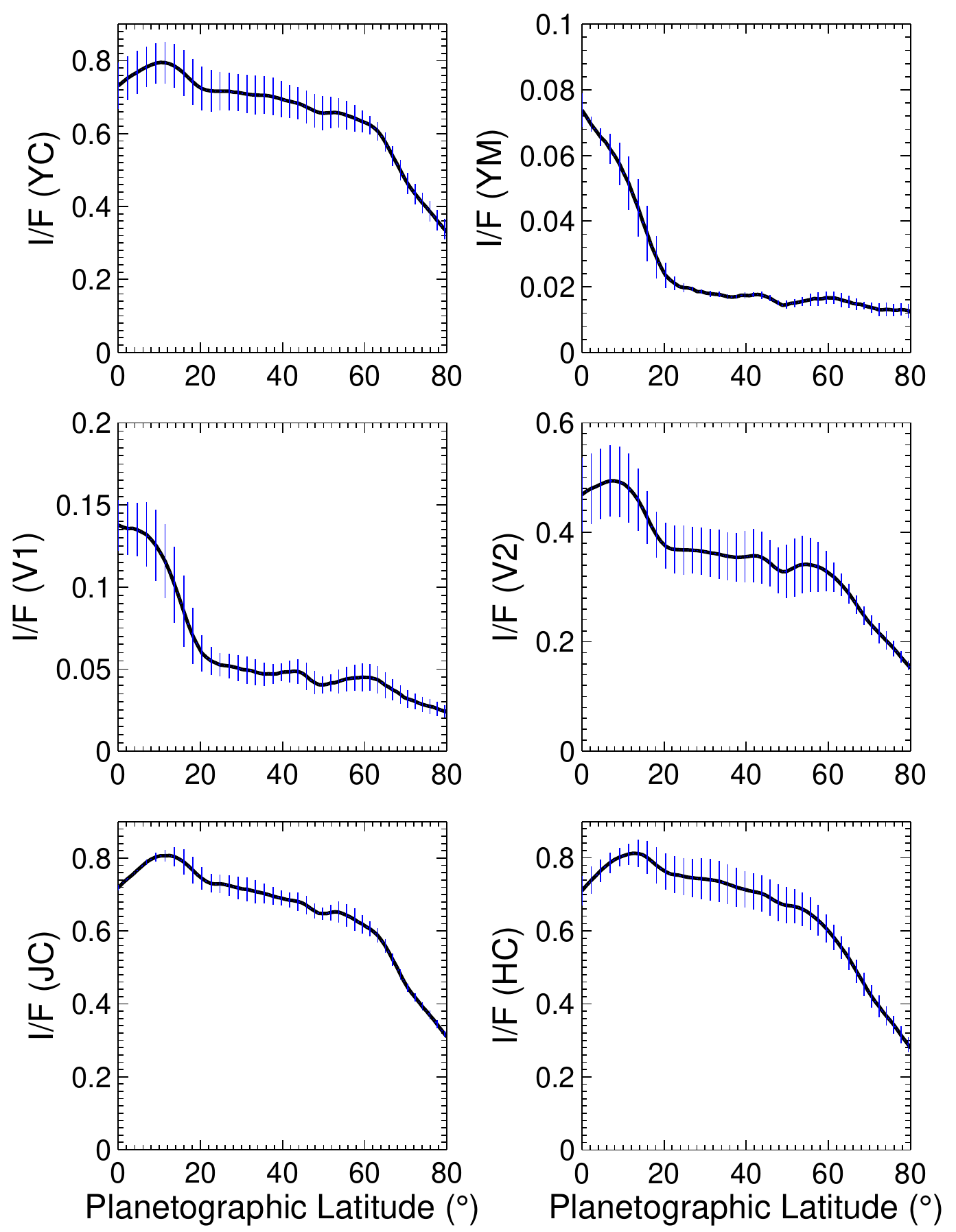}
\caption{North-south scans of mean absolute reflectivity (2015--2016) of Saturn across the central meridian in the northern hemisphere in SWIR channel filters.}
\label{satSWIRif}
\end{figure}

We show the overall mean absolute reflectivity at the central meridian of Saturn and its associated uncertainty in Figure~\ref{satVISif} for the filters in the VIS channel. The rings of Saturn cover part of the disk of the planet preventing us from observing the southern hemisphere. Latitudinal values shown here are limited to the northern hemisphere where the influence of Saturn rings is negligible. Our results correspond to observation campaigns run from April 2013 to May 2016. The number of campaigns considered and average uncertainties for these filters are as follows: UV (five; 12\%), Vio (four; 13\%), B (four; 10\%), V (four; 12\%), M1 (three; 11\%), C1 (four; 16\%), M2 (four; 9\%), C2 (three; 12\%), M3 (three; 23\%), C3 (two; 24\%).

Similar results for Saturn in the SWIR channel filters available are shown in Figure~\ref{satSWIRif} and they correspond to observations acquired between May 2015 and July 2016. The number of campaigns considered and average uncertainty for these filters are as follows: YC (four; 7\%), YM (four; 12\%), V1 (four; 17\%), V2 (three; 11\%), JC (three; 3\%), and HC (four; 6\%).

\subsection{Uncertainty sources} \label{uncertainty}

There are several sources of uncertainty and variability in the retrieved $I/F$ mean values at each latitude: the intrinsic uncertainty in the calibration procedure itself, uncertainty in the determination of the latitudes of each pixel due to small inaccuracies in the planet navigation, variations of the reflectivity at different longitudes on particular bands and, finally, possible temporal changes in the reflectivity of some planet bands observed on different epochs.

The uncertainty on $I/F$ produced by the image calibration procedure can be estimated from the errors in the different parameters that appear in Equation 1. In previous sections we estimated the calibration values dispersion up to around 20\%. Comparing count rates for a standard star with the same telescope and optics at different observation nights but similar airmass, we can find variations of 1\% to 10\% and, rarely we can find variations up to 20\%. While most of the error can be attributed to the photometrical conditions along the observing run for the planet and calibration star, other sources include detector noise and  uncertainties in the knowledge of the spectrum of the standard star used in each case; detector noise is particularly important for some filters in deep methane
absorption bands in the SWIR channel resulting in very low signals, such
as the HM filter.

Fluctuations in flat-fielding and dark current subtraction can affect count rates for the planet and standard stars. However, these variations are well below those due to atmospheric conditions of the Earth (i.e., transparency and seeing). However, such instrumental sources of uncertainties rarely account for more than 2\% of the signal \citep{mendikoa2016}. Extinction coefficients at Calar Alto Observatory are regularly monitored \citep{sanchez2007} and they imply a relative error of around 5\%. In order to minimize errors from atmospheric extinction, the standard stars were usually observed at elevations with differences below 0.3 air masses to the elevation of the planet.

Spectral data available for the standard stars also contain a certain degree of inaccuracy that contributes to the total calibration error. In particular, for some standard stars their calibrated flux was only available at low spectral resolution for some wavelength ranges. This could imply significant calibration errors for narrow filters for which spectral bands are not properly resolved in the spectrum of the standard star. As an example, for star HD19445 different spectrum sources were compared (i.e., Isaac Newton Group, SIMBAD, and X-Shooter\footnote{http://xsl.u-strasbg.fr/}) including some synthetic models based on Kurucz \citep{castelli1997} and BT-Nextgen \citep{allard2012}. For visible filters, the total star flux showed a variation of around 15\% depending on the stellar parameters assumed. For this reason, it would be advisable to promote a calibration campaign with simultaneous medium- to high-resolution spectra taken with another instrument for some selected stars.

Uncertainties introduced from inaccuracies in the planet navigation can be estimated after examining results with different navigations of the same planetary image. Our tests show that the uncertainty of a band width can be around 10\% and the latitudinal uncertainty in the position of the maximum brightness of a particular band can be around 1$^{\circ}$. These are typical values for both VIS and SWIR unprocessed photometric images but of course processed images acquired under very good seeing conditions ($<$ 0.7 arcsec) largely minimize these errors.

The different $I/F$ scans always correspond to the central meridian, but, because of the rotation of the planet, every scan corresponds to a different planetary longitude, and given that planetary bands have fluctuations in their albedo edges and that they may include discrete features (spots), the band width and reflectivity may vary. As in the disk navigation issue, we estimated uncertainty in band widths due to these variations to be $\sim$ 10\% and positions in around 1$^{\circ}$, by analyzing several $I/F$ versus latitude scans at different longitudes in the same navigated image.

Finally, variations in $I/F$ scans can also be due to intrinsic changes of the planet reflectivity over time at particular bands in Jupiter \citep{sph2012} or over large regions in Saturn as seasonal variations proceed \citep{smith1982,asl1993}. Characterizing real global or local color variations in Jupiter and Saturn clouds is a key scientific area and relevant results from PlanetCam are discussed in section~\ref{temporal} after we appropriately constrained artificial sources of variations and uncertainties.

In summary, these uncertainty sources contribute to an overall systematic uncertainty that has been estimated to be around 10-20\% in absolute reflectivity and much lower (possibly below 2\%) in relative photometry. Not perfectly photometric conditions are the main cause responsible for this level of uncertainty, while the contribution of other sources, while having a minor impact, cannot be fully discarded.

\section{Discussion}

\subsection{Latitudinal reflectivity dependence}

\begin{figure}
\centering
\includegraphics[width=8cm]{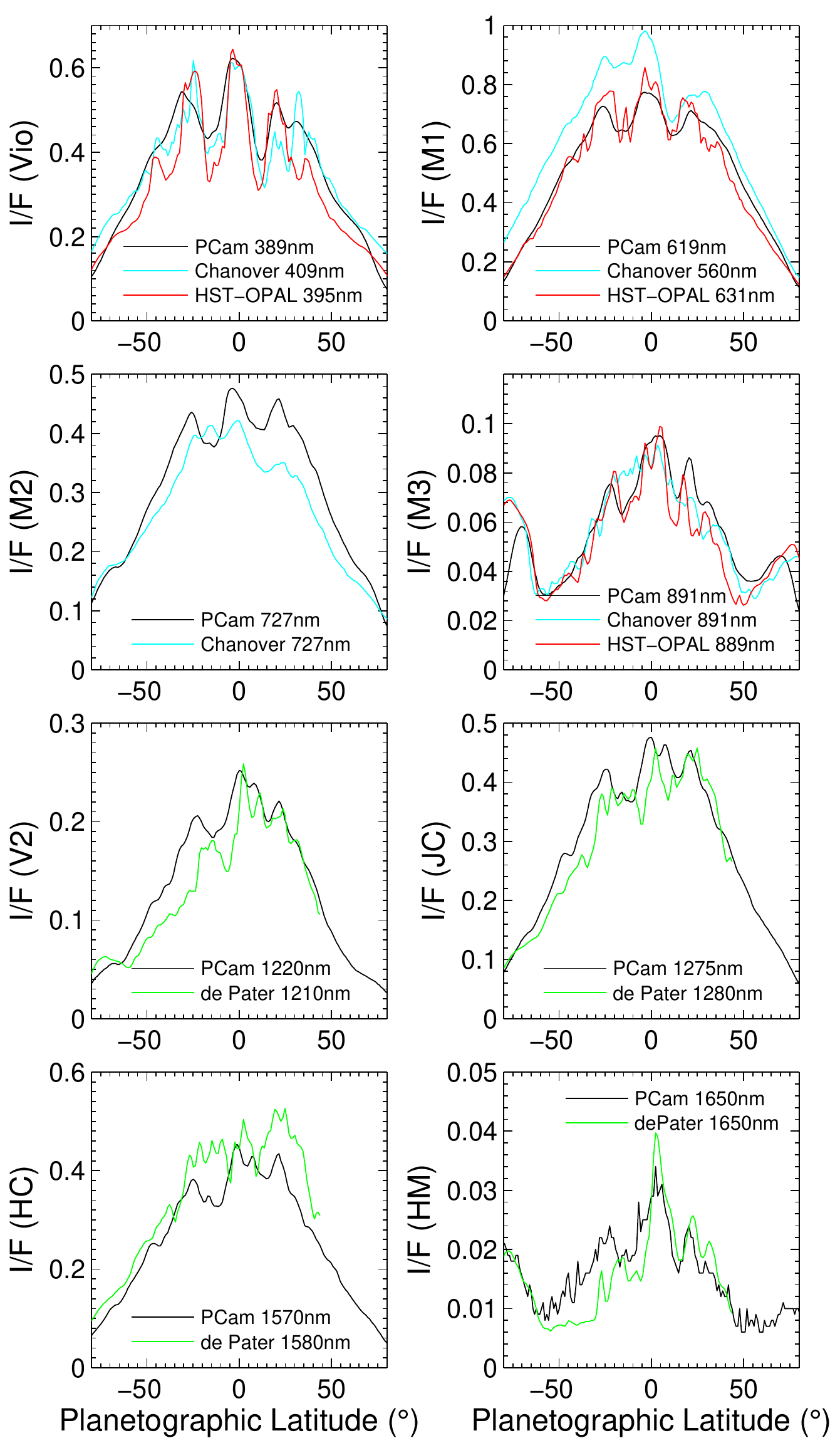}
\caption{Jupiter mean absolute reflectivity north-south scans from PlanetCam VIS and SWIR selected wavelengths compared to similar reference values from \cite{chanover1996}, HST-OPAL \citep{simon2015}, and \cite{depater2010}.}
\label{jupNSrefs}
\end{figure}

\begin{figure}
\centering
\includegraphics[width=8cm]{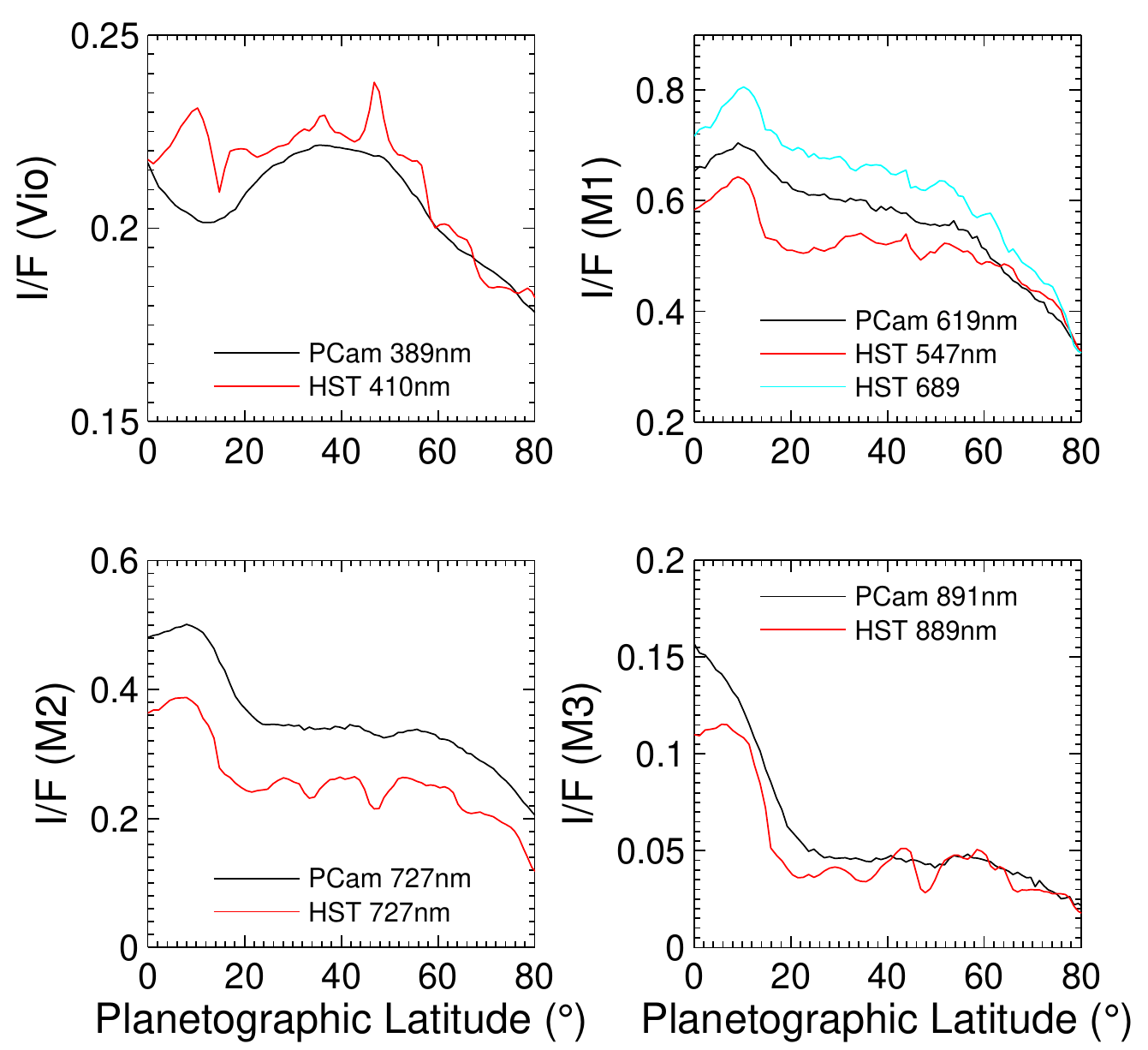}
\caption{Saturn mean absolute reflectivity in the northern hemisphere at selected wavelengths from PlanetCam VIS channel compared to reference values obtained in near spectral wavelength filters using HST images.}
\label{satNSrefs}
\end{figure}

The mean Jupiter reflectivity central meridian scans (Figures~\ref{jupVISif} and~\ref{jupSWIRif}) have been compared to available reference values from the literature \citep[][see Fig.~\ref{jupNSrefs}]{chanover1996,depater2010}. Ground-based calibrated observations presented by \cite{chanover1996} and HST calibrated images obtained in February 2016 as part of the Outer Planets Atmospheres Legacy (OPAL) program \citep{simon2015} constitute excellent sources of comparison for our Jupiter data in the visible. For the SWIR spectral range there are very few reference values of the reflectivity of Jupiter available in the literature. We compared our $I/F$ scans to values obtained by \cite{depater2010} that used filters comparable to those used in PlanetCam SWIR. When there was no perfect match between filters wavelength, we used the closest match. The differences in reflectivity between our data and the reference values may be caused by real changes in the cloud structure or in the optical properties of haze particles but part of these differences could be due to the calibration uncertainty. In general, differences above a 10\% uncertainty can be assigned to real changes in the hazes and upper Jovian cloud as in particular when compared to the reflectivity scans obtained in July 1994 by \cite{chanover1996} and our PlanetCam data obtained about 20 years later.

Regarding Saturn, HST observations obtained in July 2015 are used as reference. Figure~\ref{satNSrefs} shows a comparison of the central meridian scans corresponding to the nearest filters to PlanetCam from mean scans from April 2013 to May 2016. Significant differences can be seen in M2, where the effective wavelength and FWHM are slightly different in PlanetCam (727.3nm, 5nm) with respect to that of HST (727nm, 7.3nm). This can cause reflectivity differences in such narrowbands a M2. By contrast, the M3 band is wider and therefore more insensitive to small filters differences between PlanetCam (890.8nm, 5nm) and HST (889nm, 8.9nm).

\subsection{Albedo}
\subsubsection{Jupiter}

\begin{figure}
\centering
\includegraphics[width=8cm]{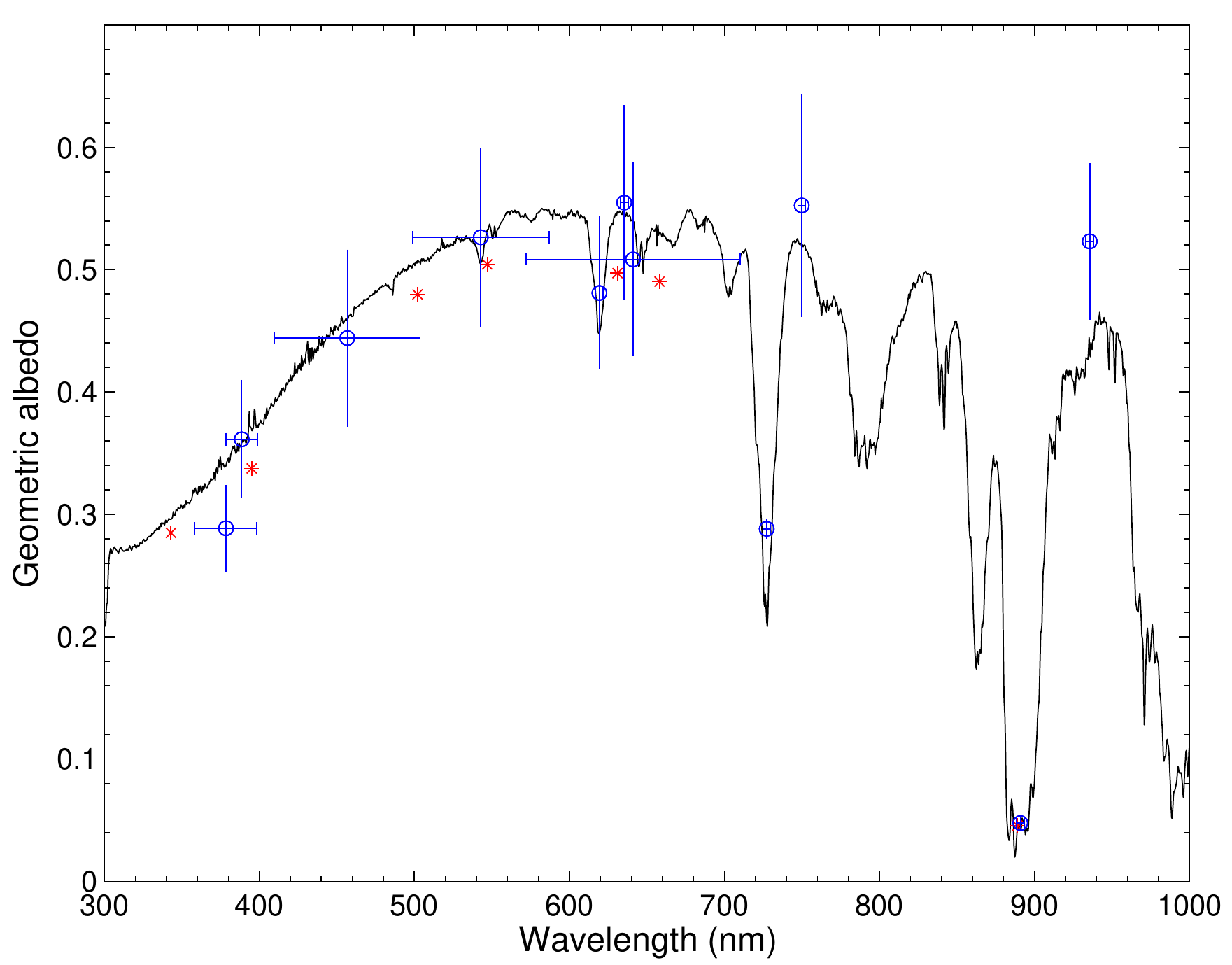}
\caption{Jupiter geometric albedo from PlanetCam VIS observations (blue dots) compared to reference values from ground-based observations in July 1995 \citep{karkoschka1998} (black line) and February 2016 HST-OPAL data \citep{simon2015} (red dots). Horizontal error bars account for the width (FWHM) of each filter, while vertical error bars indicate the uncertainty in the photometric calibration.}
\label{jupVISalbedo}
\end{figure}

\begin{figure}
\centering
\includegraphics[width=8cm]{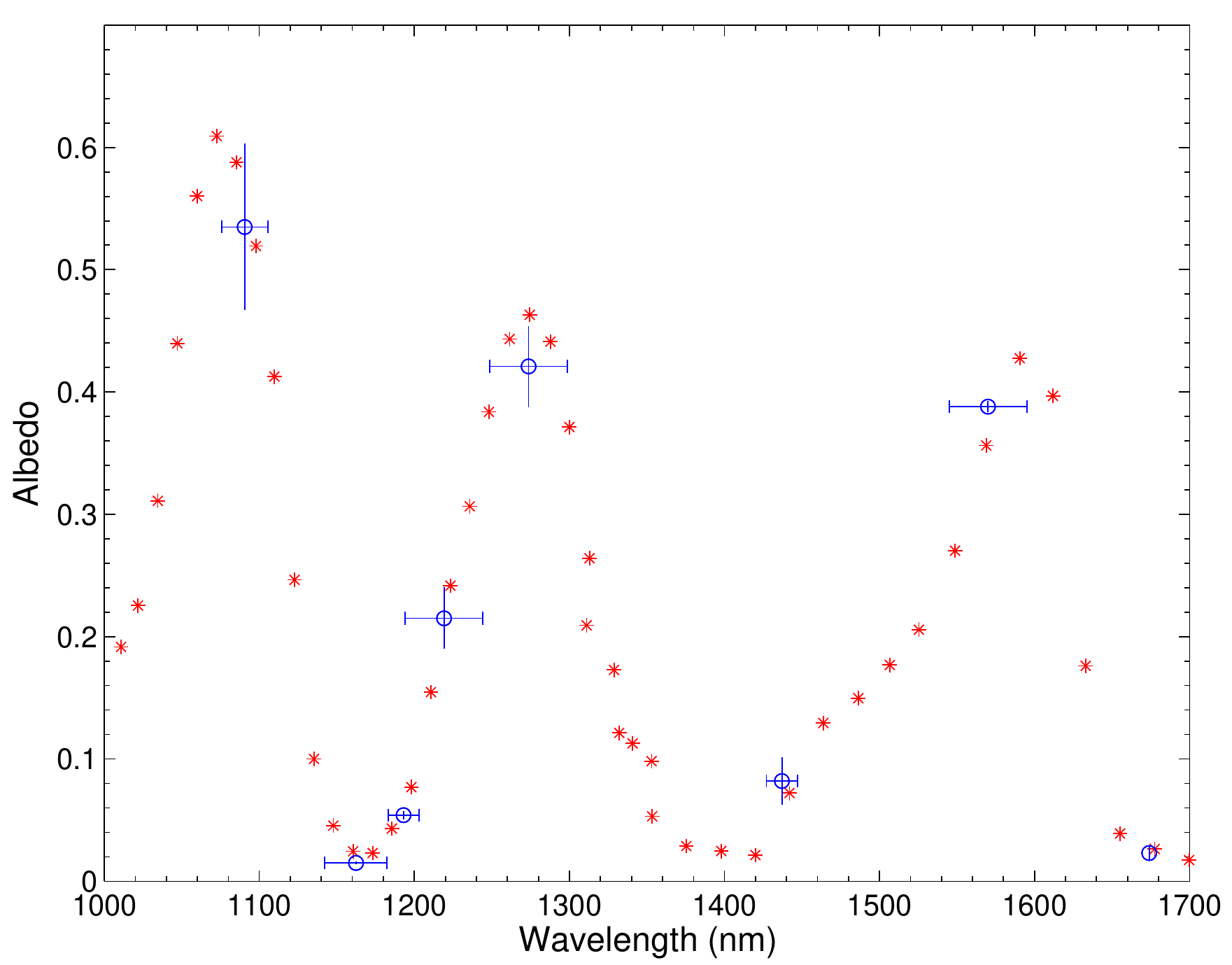}
\caption{Jupiter albedo (disk center, at latitudes $\pm$20$^\circ$) from PlanetCam SWIR observations (blue dots) compared to reference data of November 1976 \citep{clark1979} (red dots) values. PlanetCam albedo in filter YM-1160 is 0.014 \citep[0.024 in][]{clark1979}.}
\label{jupSWIRalbedo}
\end{figure}

The full disk geometric albedo for the visual spectral range is shown in Figure~\ref{jupVISalbedo} and is compared to reference values from July 1995 \citep[ESO;][]{karkoschka1998} and February 2016 \citep[HST-OPAL;][]{simon2015}. The agreement is good for most cases, although small deviations are found at the wavelengths of the filters; i.e., U, M2, and C3.

Jupiter albedo in the SWIR spectral range is shown in Figure~\ref{jupSWIRalbedo}. In this case, only the central part of the disk has been considered (reflectivity from latitudes -20$^{\circ}$ to +20$^{\circ}$) to compare our results with those from \cite{clark1979}. Again a reasonable agreement is found between both sets of values.

\subsubsection{Saturn}

\begin{figure}
\centering
\includegraphics[width=8cm]{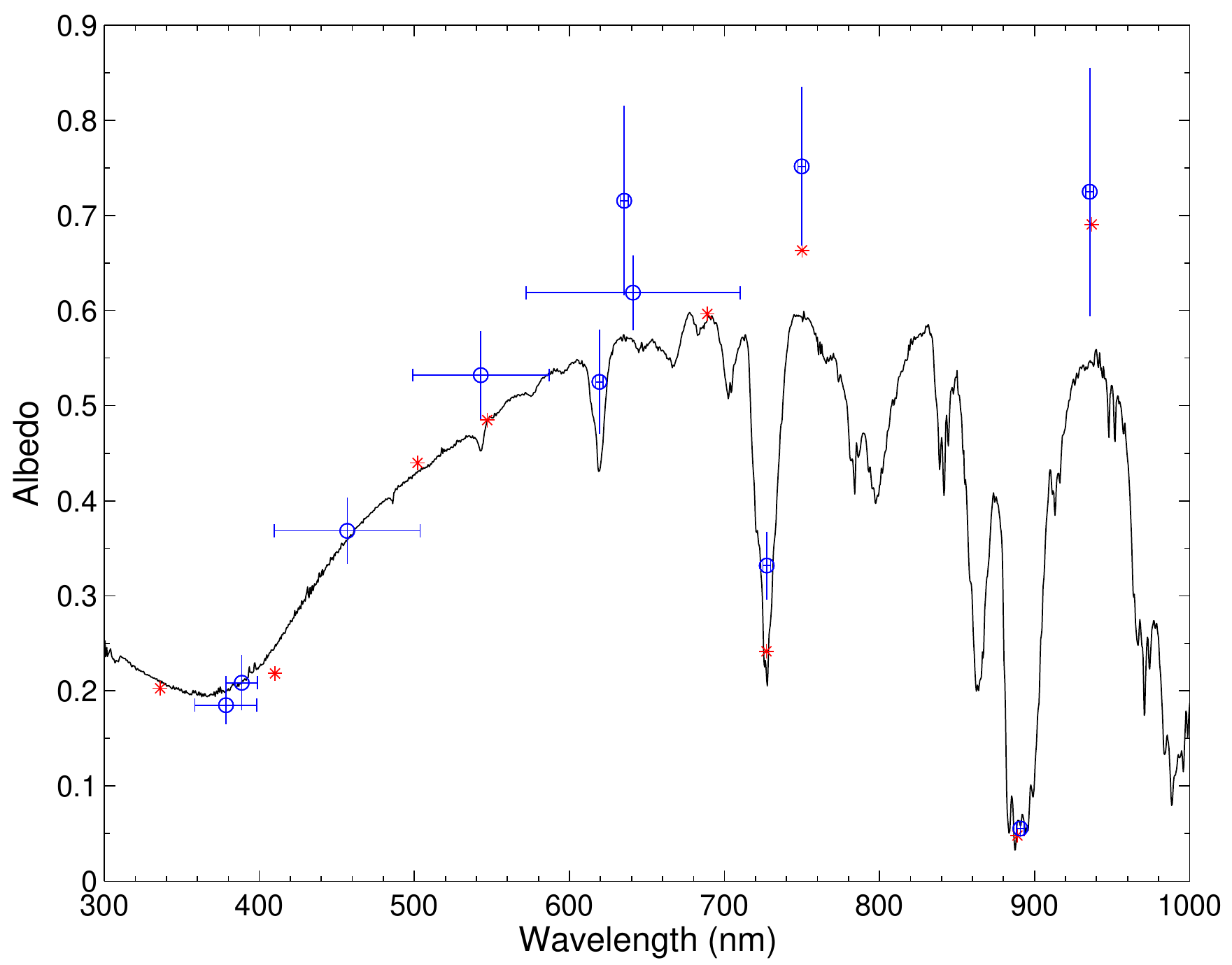}
\caption{Saturn albedo from PlanetCam VIS observations (blue dots) compared to reference results of July 1995 \citep{karkoschka2005} (black line) and HST in July 2015 \citep{asl2016a} (red dots).}
\label{satVISalbedo}
\end{figure}

\begin{figure}
\centering
\includegraphics[width=8cm]{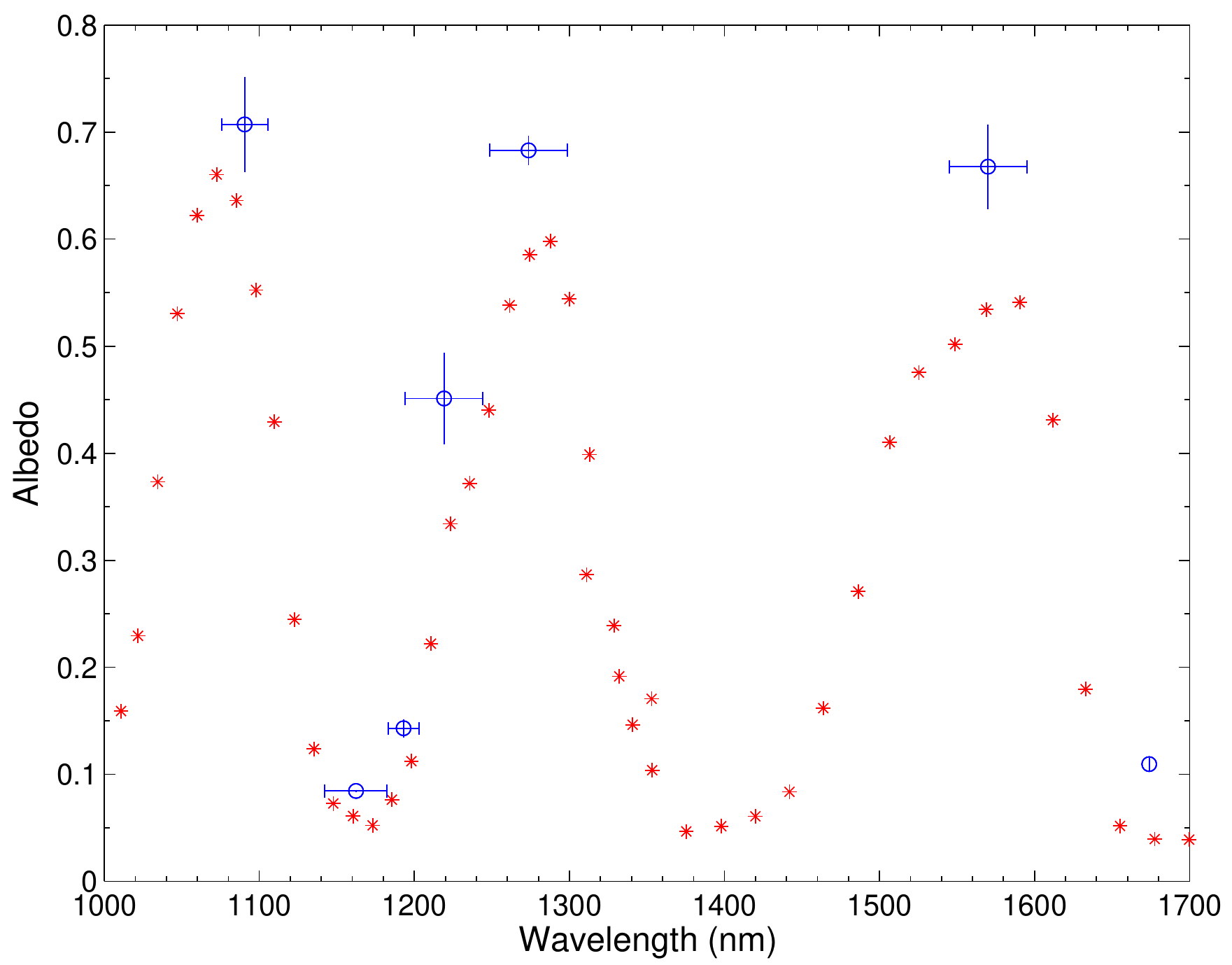}
\caption{Saturn albedo from PlanetCam SWIR observations (blue dots) compared to reference data of February 1977 \citep{clark1979} values (red dots).}
\label{satSWIRalbedo}
\end{figure}

Saturn overall mean albedo from PlanetCam campaigns is shown in Figure~\ref{satVISalbedo} and is compared to ground-based data obtained in July 1995 \citep{karkoschka1998} and recent HST July 2015 observations \citep{asl2016a}. Because of the viewing angle geometry only the ring-free northern hemisphere can be studied (see Figure 5 and 6) in both PlanetCam and HST albedo values. Our results are comparable to those retrieved from HST photometry.  There are, however, some differences when compared to full disk geometric albedo values from \cite{karkoschka1998}, which were obtained at Saturn's equinox showing intrinsic brightness differences produced by seasonal changes in the hazes and clouds \citep{sph2005} and by the different viewing or illumination geometry.

Similarly to Jupiter, the SWIR albedo measurements of Saturn were only calculated for the central part of the disk (only latitudes from -10$^{\circ}$ to +10$^{\circ}$ due to the ring presence) and were compared to data from \cite{clark1979} (Figure~\ref{satSWIRalbedo}), where a larger central disk was considered because of the different ring tilt angle. This time the comparison is closer than in the VIS case, given the similar seasonal situation of the planet at both epochs since the subsolar latitude of Saturn was around -17$^{\circ}$ at the time of \citet{clark1979} observations in 1977 and between +17$^{\circ}$ and +25$^{\circ}$ for PlanetCam observations (2012-2016), while it was around 0$^{\circ}$ by the time of Karkoschka observations in 1995.

\subsection{Temporal evolution} \label{temporal}

The temporal evolution of the properties of the cloud cover of Jupiter and Saturn can be analyzed from the absolute reflectivity $I/F$ scans comparing results obtained in different observing campaigns. A key issue in this analysis is to estimate the uncertainty values from the different sources as described in section~\ref{uncertainty}, so that the observed variations can be clearly attributed to physical atmospheric variations over time.

In order to minimize the systematic uncertainty in absolute reflectivity calibration, we scaled the different $I/F$ scans by a ratio given by the corresponding planet albedo to the overall mean albedo. Thus, we assumed that full disk geometric albedo is constant over the four years of PlanetCam observations and we scaled $I/F$ curves according to this assumption. As long as this assumption is valid and the other uncertainty sources related to image navigation (particularly remarkable at high latitudes) and planet bands width variations along the whole longitudes are known, the remaining variations between the different observing campaigns can be interpreted as caused by temporal changes in the cloud cover of these planets.

\subsubsection{Jupiter}

\begin{figure}
\centering
\includegraphics[width=8cm]{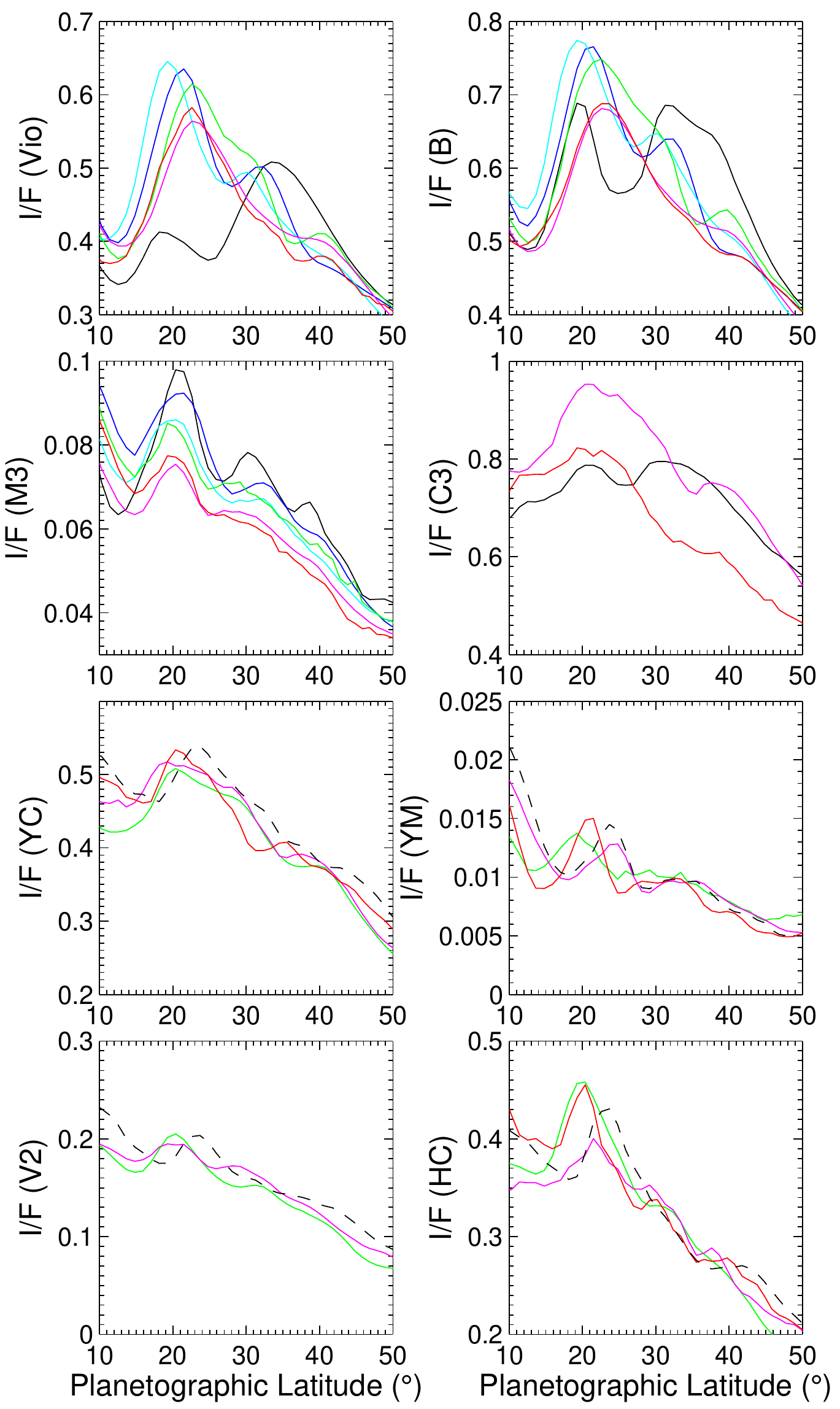}
\caption{Absolute reflectivity of Jupiter at selected wavelengths from December 2012 to July 2016. The black line corresponds to December 2012, blue to December 2013, cyan to April 2014, green to March 2015, magenta to March 2016, red to May2016 and the black dashed line to July 2016.}
\label{jupchanges}
\end{figure}

Figure~\ref{jupchanges} shows the Jupiter central meridian reflectivity evolution of the belts and zones from December 2012 to May 2016. Some changes are remarkable for example in the northern hemisphere. In the violet (Vio) filter we observe at 20$^{\circ}$ planetographic latitude that the reflectivity increases from a peak $I/F$ of 0.41 in December 2012 to 0.64 in April 2014, decreasing down to 0.56 in March 2016. We can also compare these $I/F$ peak values to the averaged reference value of 0.55, so a variation of up to 40\% can be observed in this spectral band.  We estimate that the uncertainties due to the error sources above discussed can be on the order of 10\%.

\begin{figure*}
\centering
\includegraphics[width=17cm]{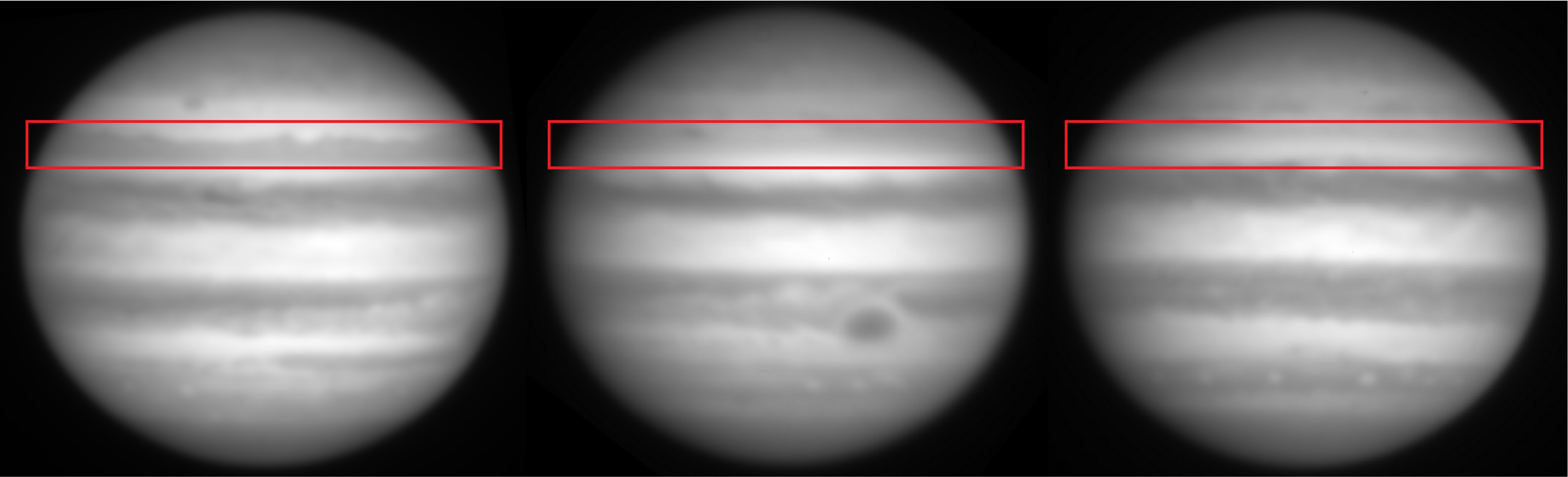}
\caption{Temporal changes in the albedo of a dark band at latitude 25$^{\circ}$ in Jupiter at the B filter. From left to right, December 2012 (sub-Earth latitude +3.01$^\circ$), April 2014 (sub-Earth latitude +1.56$^\circ$), and May 2016 (sub-Earth latitude -1.54$^\circ$) are shown.}
\label{jupchanges_ex}
\end{figure*}

Similar variations can be observed in the blue (B) filter, where $I/F$ peak at latitude +20$^{\circ}$ varies from 0.69 in Dec 2012 (0.98 times the average of this peak in B filter) to 0.77 in April 2014 (1.1 is the global average) and back to 0.68 in March 2016 (0.97 is the global average).  A dark belt at planetographic latitude +25$^{\circ}$ present in December 2012 vanishes in the following years (Figure~\ref{jupchanges_ex}). This corresponds to the evolution of an eruption in the NTB in 2012 close to solar conjunction that could not be observed properly. A similar phenomena was observed in 2007 \citep{asl2008} and 2017 \citep{asl2017}.

The reflectivity in the methane band filters show the opposite behavior at these latitudes. The $I/F$ scans in M3 show a peak at latitude +20$^{\circ}$ continuously decreasing from 0.098 in Dec 2012 (1.14 times the average) to 0.086 in April 2014 (1.00 times the average) and 0.075 in March 2016 (0.87 the average), i.e., a total reflectivity decrease of 27\% from the average. Smaller variations can be seen in the 2$^{\circ}$ wide band at latitude around 45º with a variation of 10\% with respect to the four-year average at that latitude. The nearby continuum filter C3 does not show any of these reflectivity values decrease, suggesting altitude variations in the top of a haze layer. On the other hand, variations over the time at this latitude are also seen in the continuum HC filter changing from 0.46 in March 2015 (1.09 times the average) to 0.40 in March 2016 (0.95 the average). Ongoing analyses of these data and their interpretation in terms of radiative transfer models will be presented elsewhere.

\subsubsection{Saturn}

\begin{figure}
\centering
\includegraphics[width=8cm]{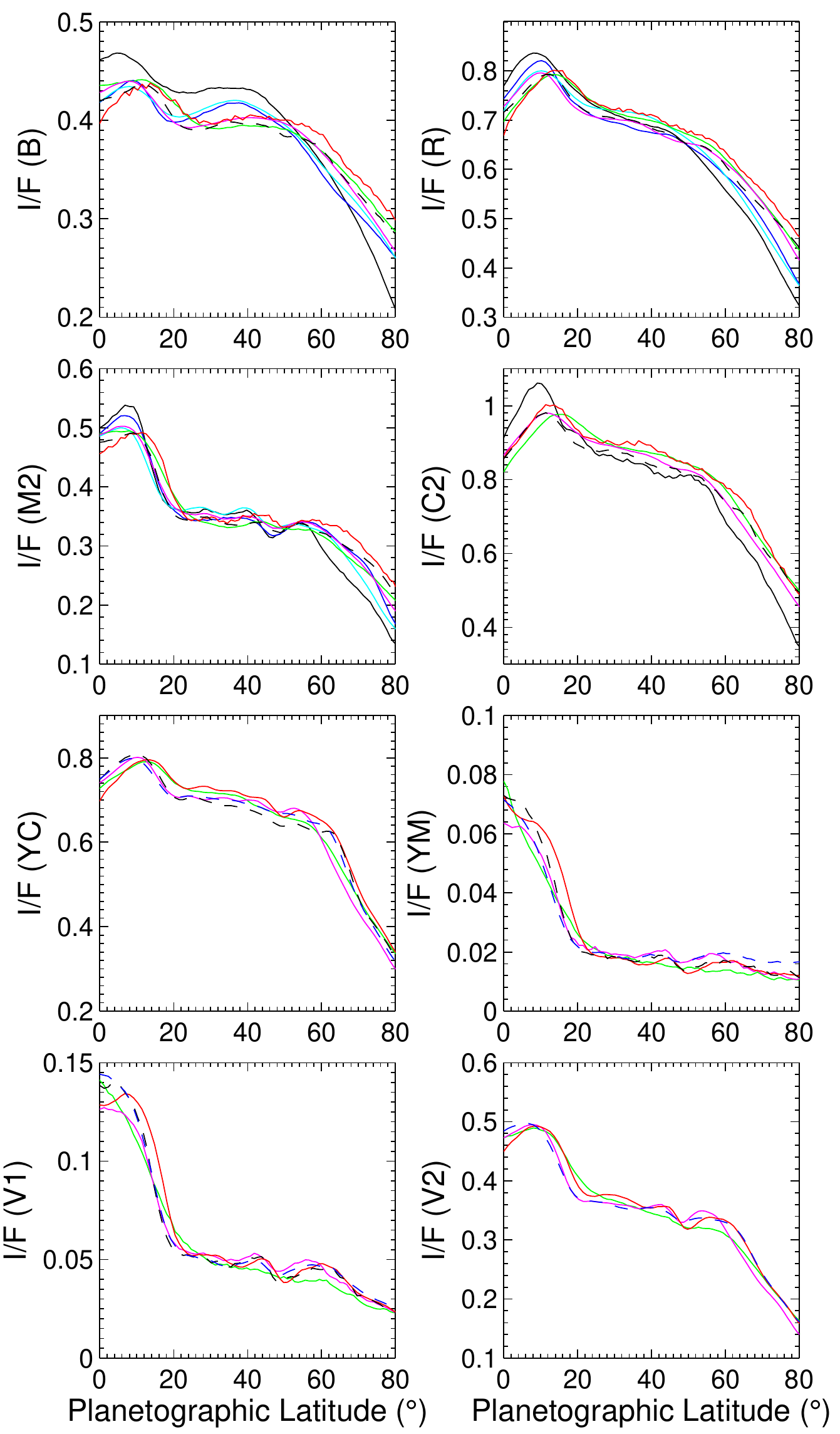}
\caption{Temporal absolute reflectivity north-south scans of the northern hemisphere of Saturn at selected wavelengths for the period from April 2013 to May 2016. The black line corresponds to April 2013, blue to April 2014, cyan to July 2014,  green to May 2015,  magenta to July 2015, red to March 2016, and the black dashed line  to May2016 and blue dashed to July 2016.}
\label{satchanges}
\end{figure}

In Figure~\ref{satchanges} we show the spectral reflectivity scans for the ring-free northern hemisphere under the viewing angle geometry of the period 2013-2016. We have to make similar considerations as described in the case of Jupiter, regarding the different possible sources of the yearly averaged variations. The most remarkable differences we detect between these campaigns occur between the brightest part near the equator and the polar zones. In addition, smaller variations in some specific belts and zones can be observed.

Globally, the northern hemisphere of Saturn showed variations of its temporal reflectivity along the observing period (4 years), in particular at visual wavelengths, which is in agreement with previous observations \citep{sph2006}. Temporal variability at different latitudes is detected in particular in the blue filter; this is probably related to changes in the upper aerosol haze layers (stratospheric and tropospheric) sensitive to short wavelength absorption \citep{sph2005,sph2016}. The equatorial zone (0$^{\circ}$ to +20$^{\circ}$) and the polar area (+60$^{\circ}$ to +90$^{\circ}$) are the regions in which most of the reflectivity temporal variability is seen. Changes in the blue and red filters can reach up to 30\% relative to the mean values. Curiously, the variability is much less pronounced in the SWIR channel, while changes around 5\% can also be detected for example at the filter V2. A study of the origin of this variability will be performed and presented elsewhere.

\section{Summary and conclusions}

The main conclusions from our study are summarized in the following points:

\begin{enumerate}
        \item We presented spatially resolved absolute reflectivity measurements of Jupiter and Saturn in a wide wavelength range at specific filters of planetary interest for the study of the cloud structures of both planets from the ultraviolet (380 nm) to short near-infrared wavelengths (1.7 $\mu$m).

        \item We performed photometry with a lucky-imaging camera, PlanetCam-UPV/EHU, specifically adapted for the study of solar system objects at high spatial resolution. Our photometrically calibrated images, without any further processing, reaches spatial resolution of up to 0.5'', with typical values of around 1'' for narrow filters.

        \item We determined the absolute reflectivity $I/F$ with a wide range of standard stars. Although the precision of these measurements depends on wavelength, we typically found that a precision of 10-20\% is achieved in the determination of absolute $I/F$ values.

        \item We presented Minnaert coefficients, which are useful to characterize the reflectivity of giant planet atmospheres under different viewing angles.

        \item We validated the photometric calibration through a comparison with published data available in the literature. Photometrically calibrated images or albedo values of giant planets are scarce in the scientific literature and the capability of determining the absolute reflectivity of Jupiter and Saturn in different years using the lucky-imaging method can contribute relevant data to this field.

        \item The $I/F$ north-south scans along the central meridian of Jupiter and Saturn for each spectral filter show temporal changes in the upper aerosol haze layers along the period studied (2012-2016).

        \item The high spatial resolution provided by the lucky-imaging method allows us to follow the spectrally dependent reflectivity changes in belts and zones, but also in specific features like the Great Red Spot, oval BA, and other cyclones, anticyclones, and storms in Jupiter and Saturn.

        \item PlanetCam-UPV/EHU is providing a long baseline of absolutely calibrated observations of the giant planets with at least two to three observation campaigns each year.

        \item We are able to detect relative changes in reflectivity of around 5-10\% when assuming the invariability of the geometric albedo of the planet at a given wavelength.

        \end{enumerate}

Some of the results we have presented will be further analyzed in future publications in terms of a radiative transfer model to retrieve information related to the vertical structure of the cloud systems of Jupiter and Saturn. Our database of planetary calibrated images, such as those presented here, is expected to be enlarged with additional observations in the future. We expect that PlanetCam observations will be able to support space mission such as Akatsuki at Venus \citep{asl2016b}, Mars Reconnaissance Orbiter, Mars Express and Exomars at Mars \citep{asl2015}, Juno at Jupiter \citep{hueso2017b,asl2017}, and to help the study of the legacy of the Cassini mission. Observations in the SWIR channel (1-1.7 microns) might be useful  when comparing with future JWST observations starting in 2019. The long-term observations will serve to cover the study of those planets where no space missions are expected in the near future as for Saturn, after the Cassini grand finale in September 2017, along with Uranus and Neptune.

\begin{acknowledgements}
This work was supported by the research project AYA2015-65041-P (MINECO/FEDER, UE), Grupos Gobierno Vasco IT-765-13, UPV/EHU UFI11/55 and ''Infraestructura'' grants from G.Vasco and UPV/EHU. I. Mendikoa has been supported by Aula Espazio Gela funded by Diputación Foral de Bizkaia. J.L.-S. acknowledges the Office of Naval Research Global (award no. N62909-15-1-2011) for support. This research has made use of the VizieR catalog access tool, CDS, Strasbourg, France. The original description of the VizieR service was published in A\&AS 143, 23.
\end{acknowledgements}

%
%

\bibliographystyle{aa} 
\bibliography{31109_am_IM} 

\end{document}